\def\year{2018}\relax
\documentclass[letterpaper]{article} 
\usepackage{aaai18}  
\usepackage{times}  
\usepackage{helvet}  
\usepackage{courier}  
\usepackage{url}  
\usepackage{graphicx}  
\usepackage{stmaryrd} 
\usepackage{epsfig,graphicx} 
\usepackage{amsthm,amsmath,amssymb}
\usepackage{bbm}
\usepackage{latexsym}
\usepackage[norelsize,ruled,vlined,linesnumbered]{algorithm2e}
\usepackage{tikz}
\usepackage{mathrsfs}
\usepackage{hyperref}
\usetikzlibrary{patterns}
\usetikzlibrary{decorations.pathreplacing}

\long\def\comment#1{}

\newtheorem{definition}{Definition}
\newtheorem{theorem}{Theorem}

\newtheorem{lemma}{Lemma}

\newtheorem{example}{Example}
\newenvironment{proofsketch}{\noindent \textit{Proof}\textit{ (Sketch)}.}{$\Box$\\ }
\newenvironment{proofsketch2}{\noindent \textit{Proof}\textit{ (Sketch)}.}{}

\newenvironment{theorem-appendix}{\vspace*{0.25cm}\textit{Theorem} }{} 
\newenvironment{proposition-appendix}{\vspace*{0.25cm}\textit{Proposition} }{} 
\newenvironment{lemma-appendix}{\vspace*{0.25cm}\textit{Lemma} }{} 

\newcommand{\vect}{\mathbf}

\newcommand{\SC}{\textsc}
\newcommand{\SF}{\textsf}

\newcommand{\LA}{\langle}
\newcommand{\RA}{\rangle}
\newcommand{\ra}{\rightarrow}

\newcommand{\REST}{\hspace{-0.05in}\restriction}

\newcommand{\CONST}{\textsc{const}}
\newcommand{\TERMS}{\textsc{terms}}
\newcommand{\NULLS}{\textsc{nulls}}
\newcommand{\VAR}{\textsc{var}}
\newcommand{\UVAR}{\forall\mbox{-}\textsc{var}}
\newcommand{\EVAR}{\exists\mbox{-}\textsc{var}}
\newcommand{\MVAR}{\textsc{m-var}}
\newcommand{\ARG}{\textsc{arg}}

\newcommand{\tgd}{\sigma}
\newcommand{\tgds}{\Sigma}
\newcommand{\ATOMS}{\textsc{atoms}}

\newcommand{\REL}{\textsc{rel}}
\newcommand{\FCTGDS}{\textsc{fc-tgds}}
\newcommand{\TG}{\textsc{tg}}

\newcommand{\CYCNULL}{\textsc{cyc-null}}
\newcommand{\DOM}{\textsc{dom}}

\newcommand{\atoma}{\mathbf{a}}
\newcommand{\atomb}{\mathbf{b}}
\newcommand{\atomc}{\mathbf{c}}
\newcommand{\atomd}{\mathbf{d}}
\newcommand{\atomh}{\mathbf{h}}
\newcommand{\vatoma}{\overrightarrow{\mathbf{a}}}
\newcommand{\vatomb}{\overrightarrow{\mathbf{b}}}
\newcommand{\vatomc}{\overrightarrow{\mathbf{c}}}
\newcommand{\vatomd}{\overrightarrow{\mathbf{d}}}
\newcommand{\BD}{\textsc{bd}}
\newcommand{\NULLSET}{\textsc{nullset}}
\newcommand{\HD}{\textsc{hd}}

\newcommand{\LINK}{\textsc{link}}

\newcommand{\CHASE}{\textsf{chase}}

\newcommand{\LEVEL}{\textsc{level}}
\newcommand{\MAX}{\textsc{max}}

\title{A New Finitely Controllable Class of Tuple Generating Dependencies: 
       The Triangularly-Guarded Class}
\author{Vernon Asuncion and Yan Zhang\\
	    School of Computing, Engineering and Mathematics\\
	    Western Sydney University, Australia}

\begin{document}

\maketitle

\begin{abstract}
	In this paper we introduce a new class of tuple-generating dependencies (TGDs) 
	called {\em triangularly-guarded} (\TG) TGDs. 
	We show that conjunctive query answering under
	this new class of TGDs is decidable since this new class of TGDs also satisfies the finite controllability (FC)
	property. We further show that this new class strictly contains
	some other decidable classes such as weak-acyclic, guarded, sticky and shy.
	In this sense, the class \TG\,provides a unified representation of all these aforementioned classes of TGDs.		 
\end{abstract}

\section{Introduction}\label{intro}

In the classical \emph{database management systems} (DBMS) setting,
a query $Q$ is evaluated against a database $D$. However, it 
has come to the attention of the database community the
necessity to also include ontological reasoning and description
logics (DLs) along with standard database techniques
\cite{CalvaneseGLLR07}. As such,
the \emph{ontological database management systems} (ODBMS) has arised.  
In ODBMS, the classical 
database is enhanced with an ontology \cite{BaaderBL16} in the form
of logical assertions that generate new intensional
knowledge. An expressive form of such logical
assertions is the so-called tuple-generating
dependencies (TGDs), i.e.,
Horn rules extended by allowing 
existential quantifiers to appear in the rule heads
\cite{Cabibbo98,Patel-SchneiderH07,CaliGL09}.
 
Queries are evaluated against a database $D$ and set of TGDs $\tgds$
\big(i.e., $D$ $\cup$ $\tgds$\big) rather than just $D$, as in the classical setting.
Since for a given database $D$, a set $\tgds$ of TGDs, and a conjunctive query $Q$, the problem of determining if 
$D$ $\cup$ $\tgds$ $\models$ $Q$, i.e., the \emph{conjunctive query answering} (CQ-Ans) problem,  
is \emph{undecidable} in general \cite{BeeriV81,BagetLMS11,Rosati11,CaliGP12,CaliGK13}, a major research
effort has been put forth to identifying syntactic conditions on TGDs for which CQ-Ans is decidable.
Through these efforts, we get the decidable syntactic classes:
\emph{weakly-acyclic} (\SC{wa}) \cite{FaginKMP05}, 
\emph{acyclic graph of rule dependencies} (a\SC{grd}) \cite{BagetLMS11}, 
\emph{linear}, \emph{multi-linear}, \emph{guarded}, \emph{weakly-guarded} (\SC{w-guarded}) \cite{Rosati06,CaliGK13}, 
\emph{sticky}, \emph{sticky-join}, \emph{weakly-sticky-join} (\SC{wsj}) \cite{CaliGP12,GogaczM17},
\emph{shy} (\SC{shy}) \cite{LeoneMTV12} and 
\emph{weakly-recursive} (\SC{wr}) \cite{wr-2012}.
The \emph{weakly-recursive} class
is only defined for the \emph{simple} TGDs, which are TGDs where the variables are only allowed
to occur once in each atom and each atom do not mention constants \cite{wr-2012}.

Another research direction
that sprangs up from those previously identified classes is the possibility of obtaining more 
expressive languages by a direct combination (i.e., union) of those classes, e.g., see
\cite{KrotzschR11,CaliGP12,g13,GottlobMP13}. A major challenge then in this 
direction is that the union of two decidable classes is not necessarily decidable \cite{BagetLMS11},
e.g., it has been shown in \cite{GottlobMP13} that the union of the classes linear and sticky TGDs is 
undecidable. 

At a model theoretic level, the results in \cite{Rosati06} and \cite{BaranyGO10} 
had respectively shown that the finite controllability property holds for the linear and 
guarded fragments of TGDs. Here, a set of TGDs $\tgds$ is said to have the \emph{finite controllability} (FC)
property if for any database $D$ and conjunctive query $Q$, we have that the \emph{first-order} (FO) 
theory $D$ $\cup$ $\tgds$ $\cup$ $\{\neg Q\}$ has the finite model property. 
It is folklore that a class of FO theories ${\cal C}$ 
is said to have the \emph{finite model} (FM) property if $\phi$ $\in$ ${\cal C}$ satisfiable iff
$\phi$ has a finite model. 
The recent work in \cite{GogaczM17} has further extended the 
result in \cite{Rosati06} for linear TGDs into the sticky-join TGDs, while a more recent work by
\cite{AmendolaLM17} shows that the FC property also for the shy class \cite{LeoneMTV12}. 
As will be revealed from this paper, our work further generalizes these previous results.


Despite these efforts, there are still some examples of simple 
TGDs that do not fall under the aforementioned classes. 
\begin{example}\label{motive_examp}
	Let $\tgds_1$ be a set of TGDs comprising of the following rules:	

	\vspace*{-0.50cm}

	\begin{align}
		\tgd_{11}:&\,\,\,\SF{t}(X,Y)\wedge\ra\exists Z\,\SF{t}(Y,Z)\wedge\SF{u}(Y,Z),\label{TGD_examp_rule_1}\\
		\tgd_{12}:&\,\,\,\SF{t}(X,Y)\wedge\SF{u}(Y,Z)\ra\SF{t}(Y,Z)\wedge\SF{u}(X,Y).\label{TGD_examp_rule_2}
	\end{align}

	\vspace*{-0.1cm}

\end{example}

\noindent
Then it can be checked that $\tgds_1$ does not fall into any of the classes previously mentioned above,
and neither is it \emph{glut-guarded} (\SC{g-guarded}) \cite{KrotzschR11} nor 
\emph{tame} (\SC{tame}) \cite{GottlobMP13}.
On the other hand, because none of the head atoms ``$\SF{t}(Y,Z)$" and ``$\SF{u}(X,Y)$" of $\tgd_{12}$ mentions
the two cyclically-affected body variables ``$X$" and ``$Z$" together
(which is under some pattern that we will generalize in Section 3), 
then it can be shown that for any database $D$ and query $Q$, 
it is sufficient to only consider a finite number of \emph{labeled nulls} in $\CHASE(D,\tgds_1)$
to determine if $\CHASE(D,\tgds_1)$ $\models$ $Q$.
Actually, $\tgds_1$ falls under a new class of TGDs we call \textit{triangularly-guarded}, which 
strictly contains several of the main syntactic classes, including $\SC{wa}$, $\SC{w-guarded}$, 
$\SC{wsj}$, $\SC{g-guarded}$, $\SC{shy}$, $\SC{tame}$ and $\SC{wr}$.

The rest of the paper is structured into four main parts as follows: Section 2 provides background notions and definitions
about databases, TGDs and the problem of (boolean) conjunctive query answering; Section 3 introduces 
the triangularly-guarded ($\TG$) class of TGDs; while Section 4 presents the main results and shows 
that $\TG$ is both decidable and strictly contains some of the main syntactic classes mentioned above; Finally, Section 5
concludes the paper with some remarks.

\vspace*{-0.3cm}

\section{Preliminaries}\label{prelim}

\subsection{Basic notions and notations}\label{prelim_rel_TGDs}

We assume three countably infinite pairwise disjoint sets $\Gamma_{\cal V}$, $\Gamma_{\cal C}$ and $\Gamma_{\sf N}$ of 
\textit{variables}, \textit{constants} and \textit{labeled nulls}, respectively. 
We further assume that $\Gamma_{\cal V}$ is partitioned into two disjoint sets
$\Gamma_{\cal V}^\forall$ and $\Gamma_{\cal V}^\exists$ 
(i.e., $\Gamma_{\cal V}$ $=$ $\Gamma_{\cal V}^\forall$ $\cup$ $\Gamma_{\cal V}^\exists$), where
$\Gamma_{\cal V}^\forall$ and $\Gamma_{\cal V}^\exists$ denote the sets
of \emph{universally} ($\forall$) and \emph{existentially} ($\exists$) quantified variables, respectively.
We also assume that the set of labeled nulls $\Gamma_{\sf N}$ contains elements of the form
$\{{\sf n}_i$ $\mid$ $i$ $\in$ $\mathbb{N}\}$, where $\mathbb{N}$ is the set of natural numbers.
Intuitively, $\Gamma_{\sf N}$ is the set of ``fresh" Skolem terms that are disjoint from the set of
constants $\Gamma_{\cal C}$. 

A \textit{relational schema} ${\cal R}$ (or just \textit{schema}) is a set of 
\textit{relational symbols} (or \textit{predicates}), where each is associated with some
number $n$ $\geq$ $0$ called its \textit{arity}. We denote by $r/n$ as the relational 
symbol $r$ $\in$ ${\cal R}$ whose arity is $n$, and by $|r|$ as the arity of $r$, i.e.,
$|r|$ $=$ $n$. We further denote by $r[i]$ as the $i$-th \textit{argument} 
(or \textit{attribute}) of $r$ where $i$ $\in$ $\big\{0$,$\ldots$,$|r|\big\}$. 
We denote by $\ARG(r)$ as the set of arguments 
$\big\{a=r[i]$ $\mid$ $i$ $\in$ $\{0$,$\ldots$,$|r|\}\big\}$ of $r$
\footnote{Sometimes for convenience and when clear from the context, we refer to an argument as lower-case letters 
          (e.g., ``$a$", ``$b$", etc.) rather than ``$p[i]$".}. 
We extend this notion to the set of relational symbols ${\cal R}$, i.e.,
$\ARG({\cal R})$ $=$ $\bigcup_{r\in{\cal R}}\ARG(r)$.

A \textit{term} $t$ is any element from the set $\Gamma_{\cal V}$ $\cup$ $\Gamma_{\cal C}$ $\cup$ $\Gamma_{\sf N}$.
Then an \textit{atom} $\atoma$ is a construct of the form $r(t_1,\ldots,t_n)$ such that: 
(1) $r$ $\in$ ${\cal R}$; (2) $n$ $=$ $|r|$; and
(3) $t_i$ \big(for $i$ $\in$ $\{1,\ldots,n\}$\big) is a term. We denote tuples of atoms by 
$\vatoma$, e.g., $\vatoma$ $=$ $\atoma_1\ldots\atoma_l$, $\vatomb$ $=$ $\atoma_1\ldots\atoma_m$, 
$\vatomc$ $=$ $\atomc_1\ldots\atomc_n$, etc., and its length by $|\vatoma|$.

We denote by $\REL(\atoma)$, $\TERMS(\atoma)$,
$\VAR(\atoma)$, $\CONST(\atoma)$ and $\NULLS(\atoma)$ as the relational symbol, the set of terms, variables, constants
and labeled nulls mentioned in atom $\atoma$, respectively. We extend these notions to a set or tuples of atoms
$S$ such that  $\TERMS(S)$, $\VAR(S)$, $\CONST(S)$ and $\NULLS(S)$ denote the sets 
$\bigcup_{\atoma\in S}\TERMS(\atoma)$, $\bigcup_{\atoma\in S}\VAR(\atoma)$, $\bigcup_{\atoma\in S}\CONST(\atoma)$ 
and  $\bigcup_{\atoma\in S}\NULLS(\atoma)$, respectively. 
We say that a tuple of atoms
$\vatoma$ $=$ $\atoma_1\ldots\atoma_l$ is \emph{connected} if either: (1) $\vatoma$ is an atom (i.e., a \emph{singleton}), or 
(2) $\TERMS(\atoma_i)$ $\cap$ $\TERMS(\atoma_{i+1})$ $\neq$ $\emptyset$ holds, for each $i$ $\in$ $\{1,\ldots,l-1\}$.
More specifically, we further say that $\vatoma$ is \NULLS-\emph{connected} if either:  (1) $\vatoma$ is an atom, or 
$\NULLS(\atoma_i)$ $\cap$ $\NULLS(\atoma_{i+1})$ $\neq$ $\emptyset$ holds, for each $i$ $\in$ $\{1,\ldots,l-1\}$. 

An \emph{instance} $I$ is 
any set (can be infinite) of atoms such that $\VAR(I)$ $=$ $\emptyset$, i.e., contains no variables. 
A \emph{database} $D$ is a finite set of ground atoms 
$\VAR(D)$ $=$ $\emptyset$ and $\NULLS(D)$ $=$ $\emptyset$.

For an atom $\atoma$ $=$ $r(t_1,\ldots,t_n)$, we denote by $\ARG(\atoma)\REST_X$ (resp. $\VAR(\atoma)\REST_A$) as the set of 
arguments (resp. variables) $\big\{r[i]$ $\mid$ $i\in\{1,\ldots,n\}\big\}$ 
(resp. $\big\{X$ $\mid$ $t_i$ $=$ $X$ and $r[i]$ $\in$ $A\big\}$) but \emph{restricted} to those variables 
(resp. argument positions) mentioning $X$ (resp. from $A$).
%
%

Given two sets of terms $T_1$ and $T_2$, a \emph{homomorphism} $\theta:$ $T_1$ $\longrightarrow$ $T_2$
is a function from $T_1$ onto $T_2$ such that 
$t$ $\in$ $(T_1\cap\Gamma_{\cal C})$ implies $\theta(t)$ $=$ $t$, 
i.e., identity for the constants $\Gamma_{\cal C}$. 
Then for a given atom $\atoma$ $=$ $r(t_1,\ldots,t_n)$,
a set of terms $T$ and a homomorphism $\theta:$ $\VAR(\atoma)$ $\longrightarrow$ $T$,
a \textit{substitution of $\atoma$ under $\theta$}  \big(or just \textit{substitution} for convenience\big), 
denoted $\atoma\theta$ \big(or sometimes $\theta(\atoma)$\big), 
is the atom such that $\atoma\theta$ $=$ $r\big(\theta(t_1),\ldots,\theta(t_n)\big)$. 
We naturally extend
to conjunctions of atoms $\atoma_1$ $\wedge$ $\ldots$ $\wedge$ $\atoma_n$ so that
$\theta(\atoma_1$ $\wedge$ $\ldots$ $\wedge$ $\atoma_n)$ $=$ 
$\atoma_1\theta$ $\wedge$ $\ldots$ $\wedge$ $\atoma_n\theta$.
Given two homomorphisms $\theta_1:$ $T_1$ $\longrightarrow$ $T_2$ and $\theta_2:$ $T_2$ $\longrightarrow$ $T_3$,
we denote by $\theta_2\circ\theta_1$ as the \textit{composition} of $\theta_1$ with $\theta_2$ such that  
$\theta_2\circ\theta_1:$ $T_1$ $\longrightarrow$ $T_3$ and 
$(\theta_2\circ\theta_1)(t)$ $=$ $\theta_2\big(\theta_1(t)\big)$, for all $t$ $\in$ $T_1$.
Then lastly, given again a homomorphism $\theta:$ $T_1$ $\longrightarrow$ $T_2$ and some
set of terms $T'$ $\subseteq$ $T_1$, we denote by $\theta\REST_{T'}$ as the \textit{restriction}
of the homomorphism $\theta$ to the domain $T'$ $\subseteq$ $T_1$ such that 
$\theta\REST_{T'}:$ $T'$ $\longrightarrow$ $T_2$, and where $\theta\REST_{T'}(t)$ $=$ $\theta(t)$
$\in$ $T_2$, for each $t$ $\in$ $T'$ $\subseteq$ $T_1$. 

Given a pair of atoms $\LA\atoma_1,\atoma_2\RA$ and corresponding pair of homomorphisms 
$\theta_1:\TERMS(\atoma_1)$ $\longrightarrow$ $\Gamma_{\cal C}$ $\cup$ $\Gamma_{\cal V}$ and 
$\theta_2:\TERMS(\atoma_2)$ $\longrightarrow$ $\Gamma_{\cal C}$ $\cup$ $\Gamma_{\cal V}$,
we say that $\theta_1$ and $\theta_2$
is a \emph{most general unifier} (MGU) of the pairs $\LA\atoma_1,\atoma_2\RA$ if it \emph{unifies} 
$\atoma_1$ and $\atoma_2$ (i.e., $\atoma_1\theta_1$ $=$ $\atoma_2\theta_2$), and for any unifiers 
$\vartheta_1$ and $\vartheta_2$ of $\atoma_1$ and $\atoma_2$ (i.e., $\atoma_1\vartheta_1$ $=$ $\atoma_2\vartheta_2$),
there are unifiers $\theta'_1$ and $\theta'_2$ such that $\vartheta_1$ $=$ $\theta_1\circ\theta'_1$ 
and $\vartheta_2$ $=$ $\theta_2\circ\theta'_2$. 


\subsection{TGDs, BCQ-Ans and Chase}

A \textit{tuple generating dependency} (TGD) rule $\tgd$ of schema ${\cal R}$
is a \textit{first-order} (FO) formula of the form: 
\vspace*{-0.05cm}
\begin{align}
	\forall\vect{X}\vect{Y}\big(\Phi(\vect{X},\vect{Y})\,\ra\,\exists\vect{Z} \Psi(\vect{Y},\vect{Z})\big),\label{tgd_rule}
\end{align}
\vspace*{-0.8cm}

\noindent 
where:
\begin{itemize}
	\item $\vect{X}$ $=$ $ X_1\ldots X_k$, $\vect{Y}$ $=$ $ Y_1\ldots Y_l$ and 
	      $\vect{Z}$ $=$ $ Z_1\ldots Z_m$ are pairwise disjoint tuple of variables; 
    \item $\Phi(\vect{X},\vect{Y})$ $=$ $b_1(\vect{V}_{1})\wedge\ldots\wedge b_n(\vect{V}_{n})$ is a conjunction
          of atoms such that $\vect{V}_i$ $\subseteq$ $\vect{X}\vect{Y}$ and $b_i$ $\in$ ${\cal R}$, 
          for $i$ $\in$ $\{1$,$\ldots$,$n\}$;
    \item $\Psi(\vect{Y},\vect{Z})$ $=$ $r_1(\vect{W}_1)\wedge\ldots\wedge r_m(\vect{W}_m)$ is a conjunction 
          of atoms where $\vect{W}_i$ $\subseteq$ $\vect{Y}\vect{Z}$ and
          $r_i$ $\in$ ${\cal R}$, for $i$ $\in$ $\{1$,$\ldots$,$m\}$.
\end{itemize}
For a given TGD $\tgd$ of the form (\ref{tgd_rule}), we denote by 
$\BD(\tgd)$ as the set  of atoms $\big\{b_1(\vect{V}_{1})$,$\ldots$,$b_n(\vect{V}_{n})\big\}$, which
we also refer to as the \textit{body} of $\tgd$. Similarly, by $\HD(\tgd)$ we denote the
set of atoms $\big\{r_1(\vect{W}_1),\ldots,r_m(\vect{W}_m)\big\}$, 
which we also refer to as the \textit{head}
of $\tgd$. For convenience, when it is clear from the context, we simply drop the quantifiers in
(\ref{tgd_rule}) such that a TGD rule $\tgd$ of the form (\ref{tgd_rule}) can simply be referred to as: 
$\Phi(\vect{X},\vect{Y})$ $\ra$ $\Psi(\vect{Y},\vect{Z})$. Then, for a given set of 
TGDs $\tgds$, we denote by $\ATOMS(\tgds)$ as the set of all atoms occurring in $\tgds$
such that $\ATOMS(\tgds)$ $=$ $\bigcup_{\tgd\in\tgds}\big(\BD(\tgd)\cup \HD(\tgd)\big)$, and by
$\REL(\tgds)$ as the set of all relational symbols mentioned in $\tgds$. Then lastly,
for a given rule $\tgd$ of the form (\ref{tgd_rule}) and
atom $\atoma$ $\in$ $\HD(\tgd)$, we denote by $\UVAR(\atoma)$ 
and $\EVAR(\atoma)$ as the set of variables $\VAR(\atoma)\cap\vect{Y}$ and 
$\VAR(\atoma)\cap\vect{Z}$, respectively, i.e., the set of all the universally ($\forall$) 
and existentially ($\exists$) quantified variables of $\atoma$, respectively. 
We extend this notion to the TGD rule $\tgd$ of the form (\ref{tgd_rule}) 
so that we set $\VAR(\tgd)$ $=$ $\vect{XYZ}$, $\UVAR(\tgd)$ $=$ $\vect{XY}$ and 
$\EVAR(\tgd)$ $=$ $\vect{Z}$.

A \emph{boolean conjunctive query} (BCQ) $Q$ is a FO formula $\exists\vect{X}\varphi(\vect{X})$
$\ra$ $q$ such that $\varphi(\vect{X})$ $=$ $r_1(\vect{Y}_1)\wedge\ldots\wedge r_l(\vect{Y}_l)$,
where $r_i$ $\in$ ${\cal R}$ and $\vect{Y}_i$ $\subseteq$ $\vect{X}$, 
for each $i$ $\in$ $\{1$, $\ldots$, $l\}$, and where we set 
$\BD(Q)$ $=$ $\{r_1(\vect{Y}_1),\ldots,r_l(\vect{Y}_l)\}$. 
Given a database $D$ and a set of TGDs $\tgds$,
we say that $D$ $\cup$ $\tgds$ \emph{entails} $Q$, denoted $D$ $\cup$ $\tgds$ $\models$ $Q$,
iff $D$ $\cup$ $\tgds$ $\models$ $\exists\vect{X}\varphi(\vect{X})$. The central problem tackled 
in this work is the \emph{boolean conjunctive query answering} (BCQ-Ans): 
\emph{given a database $D$, a set of TGDs $\tgds$ and BCQ $Q$, does $D$ $\cup$ $\tgds$ $\models$ $Q$?}
It is well know that BCQ-Ans is \emph{undecidable} in general \cite{BeeriV81}.

The \emph{chase procedure} (or just \emph{chase}) 
\cite{MaierMS79,JohnsonK84,AbiteboulHV95,FaginKMP05,DeutschNR08,ZhangZY15} 
is a main algorithmic tool proposed for checking 
implication dependencies \cite{MaierMS79}. For an instance $I$, homomorphism $\eta$ and TGD 
$\tgd$ $=$ $\Phi(\vect{X},\vect{Y})$ $\ra$ $\Psi(\vect{Y},\vect{Z})$, 
we have that $I\xrightarrow{\sigma,\,\eta} I'$ defines a single \emph{chase step} as follows:
$I'$ $=$ $I$ $\cup$ $\big\{\eta'\big(\Psi(\vect{Y},\vect{Z})\big)\big\}$ such that: (1)
$\eta:$ $\vect{X}\vect{Y}$ $\longrightarrow$ $\Gamma_{\cal C}$ $\cup$ $\Gamma_{\SF{N}}$ and
$\eta\big(\Phi(\vect{X},\vect{Y})\big)$  $\subseteq$ $I$; and 
(2) $\eta':$ $\vect{X}\vect{Y}\vect{Z}$ $\longrightarrow$ $\Gamma_{\cal C}$ $\cup$ $\Gamma_{\SF{N}}$ 
and $\eta'\REST_{\vect{XY}}$ $=$ $\eta$. As in the literatures, we further assume here that
each labeled nulls used to eliminate the $\exists$-quantified variables 
in $\vect{Z}$ follows lexicographically all the previous ones, i.e.,
follows the order $\SF{n}_i$, $\SF{n}_{i+1}$, $\SF{n}_{i+2}$, $\ldots$, etc. 
A \emph{chase sequence} of a database $D$ w.r.t. to a set of TGDs $\tgds$ is a sequence of chase
steps $I_i\xrightarrow{\sigma_i,\,\eta_i} I_{i+1}$, where $i$ $\geq$ $0$, $I_0$ $=$ $D$ and
$\sigma_i$ $\in$ $\tgds$. 
An infinite chase sequence $I_i\xrightarrow{\sigma_i,\,\eta_i} I_{i+1}$
is \emph{fair} if $\eta\big(\Phi(\vect{X},\vect{Y})\big)$ $\subseteq$ $I_i$,
for some $\eta:$ $\vect{X}\vect{Y}$ $\longrightarrow$ $\Gamma_{\cal C}$ $\cup$ $\Gamma_{\SF{N}}$
and $\tgd$ $=$ $\Phi(\vect{X},\vect{Y})$ $\ra$ $\Psi(\vect{Y},\vect{Z})$ $\in$ $\tgds$,
implies
$\exists\eta':$ $\vect{X}\vect{Y}\vect{Z}$ $\longrightarrow$ $\Gamma_{\cal C}$ $\cup$ $\Gamma_{\SF{N}}$, 
where $\eta'\REST_{\vect{X}\vect{Y}}$ $=$ $\eta$, such that $\eta'\big(\Psi(\vect{Y},\vect{Z})\big)$ 
$\subseteq$ $I_k$ and $k$ $>$ $i$.  
Then finally, we let 
$\CHASE(D,\tgds)$ $=$ $\bigcup_{i=0}^{\infty}I_i$. 

\begin{theorem}\cite{DeutschNR08,CaliGP12}
	Given a database $D$, a set of TGDs $\tgds$ and BCQ $Q$, $D$ $\cup$ $\tgds$ $\models$ $Q$ iff
	$\CHASE(D,\tgds)$ $\models$ $Q$.	
\end{theorem}

\vspace*{-0.2cm}

\subsection{Cyclically-affected arguments}\label{prelim_AG}


As observed in \cite{LeoneMTV12}, the notion of \emph{affected arguments} in \cite{CaliGK13} can 
sometimes consider arguments that may not actually admit a ``firing" mapping $\forall$-variables into nulls.
For this reason, it was introduced in \cite{LeoneMTV12} the notion of a ``null-set." 
Given a set of TGDs $\tgds$,
let $\atoma$ $\in$ $\ATOMS(\tgds)$, $a$ $\in$ $\ARG(\atoma)$ and $X$ $=$ $\VAR(\atoma)\REST_{a}$. 
Then the \emph{null-set} of $a$ in $\atoma$ under $\tgds$, denoted as
$\NULLSET(a,\atoma,\tgds)$ \big(or just $\NULLSET(a,\atoma)$ if clear from the context\big), 
is defined inductively as follows: If $\atoma$ $\in$ $\HD(\tgd)$,
for some $\tgd$ $\in$ $\tgds$, then 
(1) $\NULLSET(a,\atoma)$ $=$ $\{\SF{n}_X^\tgd\}$,
\footnote{We assume that $\tgd$ $\neq$ $\tgd'$ or $X$ $\neq$ $X'$ implies 
          $\SF{n}_X^\tgd$ $\neq$ $\SF{n}_{X'}^{\tgd'}$,
          for each pair of elements $(\SF{n}_X^\tgd$, $\SF{n}_{X'}^{\tgd'})$ of all null-sets
          \cite{LeoneMTV12}.}:
if $\VAR(\atoma)\REST_{a}$ $=$ $X$ $\in$ $\EVAR(\tgd)$, or 
(2) $\NULLSET(a,\atoma)$ is the intersection of all null-sets $\NULLSET(b,\atomb)$ such that
$\atomb$ $\in$ $\BD(\tgd)$, $b$ $\in$ $\ARG(\atomb)$ and $\VAR(\atomb)\REST_{b}$ $=$ $X$.
Otherwise, if $\atoma$ $\in$ $\BD(\tgd)$, for some $\tgd$ $\in$ $\tgds$, then
$\NULLSET(a,\atoma)$ is the union of all $\NULLSET(a,\vect{a}')$ such that
$\REL(\vect{a}')$ $=$ $\REL(\vect{a})$ and $\vect{a'}$ $\in$ $\HD(\tgd')$, where
$\tgd'$ $\in$ $\tgds$.

Borrowing similar notions from \cite{KrotzschR11} used in identifying the so-called 
\emph{glut variables}, the \textit{existential dependency graph} ${\cal G}_{\exists}(\tgds)$ 
is a graph $(N,E)$, whose nodes $N$ is the union of all $\NULLSET(a,\atoma)$, where 
$\atoma\in\ATOMS(\tgds)$ and $a\in\ARG(\atoma)$, and edges:

\vspace*{-0.4cm}		

\begin{align}	
	&\hspace*{-1.3cm}E=\big\{\,\big(\SF{n}_{Z'}^{\tgd'},\SF{n}_{Z}^{\tgd}\big)\,\mid\,
	         \exists\tgd\in\tgds\mbox{ of form (\ref{tgd_rule})},\,\exists Y\in\vect{Y},\nonumber
\end{align}

\vspace*{-0.5cm}

\begin{align}	         
	      &\SF{n}_{Z'}^{\tgd'}\in\,\bigcap\NULLSET(Y,\tgd,\tgds)
	      \mbox{ and }\NULLSET(a,\atoma)=\{\SF{n}_{Z}^{\tgd}\},\nonumber
\end{align}

\vspace*{-0.5cm}

\begin{align}	      
	      &\hspace*{-3.5cm}\mbox{for some }\atoma\in\HD(\tgd)\mbox{ and }Z\in\vect{Z}\,\big\},\nonumber	
\end{align}

\vspace*{-0.1cm}

\noindent
where $\bigcap\NULLSET(Y,\tgd,\tgds)$ denotes the intersection of all $\NULLSET(b,\atomb,\tgds)$ 
such that $\atomb\in\BD(\tgd)$, $b\in\ARG(\atomb)$ and $Y=\VAR(\atomb)\,\REST_b$.
We note that our definition of a dependency graph here generalizes the \emph{existential dependency graph}
in \cite{KrotzschR11} by combining the notion of null-sets in \cite{LeoneMTV12}. 
Then with the graph ${\cal G}_{\exists}(\tgds)$ $=$ $(N,E)$ as defined above, 
we denote by $\CYCNULL(\tgds)$ as the smallest subset of $N$ such that 
$\SF{n}^{\tgd}_{Z}$ $\in$ $\CYCNULL(\tgds)$ iff either: (1) $\SF{n}^{\tgd}_{Z}$ is in a cycle in ${\cal G}_{\exists}(\tgds)$,
or (2) $\SF{n}^{\tgd}_{Z}$ is reachable from some other node $\SF{n}^{\tgd'}_{Z'}$ $\in$ $\CYCNULL(\tgds)$,
where $\SF{n}^{\tgd'}_{Z'}$ is in a cycle in ${\cal G}_{\exists}(\tgds)$.

\section{Triangularly-Guarded (TG) TGDs}

This section now introduces the triangularly-guarded class of TGDs, which is the focus of 
this paper. We begin with an instance of a BCQ-Ans problem that corresponds to 
a need of an infinite number of labeled nulls in the underlying chase derivation.

\begin{example}[\textbf{\textit{Unbounded nulls}}]\label{cyc_aag_examp}
	Let $\tgds_2$ $=$ $\{\tgd_{11},\tgd^*_{12},\}$ be the set of TGDs obtained from 
	$\tgds_1$ $=$ $\{\tgd_{11},\tgd_{12}\}$ of
	Example \ref{motive_examp} by just changing the rule $\tgd_{12}$ into the rule $\tgd^*_{12}$ such that:
	
	\vspace*{-0.5cm}	

	\begin{align}
		&\tgd^*_{12}\,=\,``\SF{t}(X,Y)\wedge\SF{u}(Y,Z)\ra\SF{t}(X,Z)\wedge\SF{u}(X,Y)".
		\label{cyc_aag_examp_form_1}
	\end{align}

	\vspace*{-0.1cm}
	
	\noindent
	Then we have that $\tgd^*_{12}$ is obtained from $\tgd_{12}$ of $\tgds_1$ by
	changing the variable ``$Y$" in the head atom ``$\SF{t}(Y,Z)$" of $\tgd_{12}$ into ``$X$", 
	i.e., to  obtain ``$\SF{t}(X,Z)$". Intuitively,
	this allows the two variables ``$X$" and ``$Y$" to act as place holders that combine labeled nulls together
	in the head atom ``$\SF{t}(X,Z)$" of $\tgd^*_{12}$.
	Now let $D_2$ $=$ $\big\{\SF{t}(c_1,c_2),\SF{u}(c_1,c_2)\big\}$ be a database, 
	where $c_1$, $c_2$ $\in$ $\Gamma_{\cal C}$ and $c_1$ $\neq$ $c_2$, and $Q_2$ the BCQ 
	$\exists X\SF{t}(X,X)\ra q$. Then 
	we have that $D_2$ $\cup$ $\tgds_2$ $\cup$ $\{\neg\exists X\SF{t}(X,X)\}$
	$\equiv$
	$D_2$ $\cup$ $\tgds_2$ $\cup$ $\{\forall X\neg\SF{t}(X,X)\}$ can only be satisfied 
	by the infinite model $M$ of the form $M$ $=$
	$D_2$ $\cup$ $\bigcup_{1\,\leq\,i\,<\,j}\big\{\SF{t}(c_i,c_j)\big\}$, where 
	we assume $i$ $\neq$ $j$ implies $c_i$ $\neq$ $c_j$. 
	Therefore, since $D_2$ $\cup$ $\tgds_2$ $\cup$ $\{\forall X\neg\SF{t}(X,X)\}$ is satisfiable
	(albeit infinitely), it follows that $D_2$ $\cup$ $\tgds_2$ $\not\models$ $Q_2$.
\end{example}

In database $D_2$ and set TGDs $\tgds_2$ $=$ $\{\tgd_{11}$, $\tgd^*_{12}\}$ of Example \ref{cyc_aag_examp}, 
we get from $\tgd_{11}$ the sequence of atoms 
$\SF{t}(c_2,\SF{n}_1)$, $\SF{t}(\SF{n}_1,\SF{n}_2)$, $\SF{t}(\SF{n}_2,\SF{n}_3)$, 
$\ldots$ $\SF{t}(\SF{n}_{k-1},\SF{n}_k)$ $\in$ $\CHASE(D_2,\tgds_2)$, where $\SF{n}_i$ $\in$ $\Gamma_{\SF{N}}$,
for each $i$ $\in$ $\{1,\ldots,k\}$. Moreover, by the repeated applications of
$\tgd^*_{12}$, we further get that $\SF{t}(\SF{n}_i,\SF{n}_k)$ $\in$ $\CHASE(D_2,\tgds_2)$,
for each $i$ $\in$ $\{1,\ldots,k-1\}$, i.e., $\SF{n}_i$ and $\SF{n}_k$, for each $i$ $\in$ $\{1,\ldots,k-1\}$, 
which will be ``pulled" together in some relation of $\SF{t}$ in $\CHASE(D_2,\tgds_2)$.
As such, for the given BCQ $Q_2$ $=$ $\exists X\SF{t}(X,X)\ra q$ also from Example \ref{cyc_aag_examp},
since $D_2\cup\tgds_2$ $\models$ $Q_2$ iff $D_2\cup\tgds_2\cup\{\forall X\neg\SF{t}(X,X)\}$
is not satisfiable, then the fact that we have to satisfy the literal ``$\neg\SF{t}(X,X)$" for all ``$X$",
and because $\SF{t}(\SF{n}_i,\SF{n}_k)$ $\in$ $\CHASE(D_2,\tgds_2)$, for $i$ $\in$ $\{1,\ldots,k-1\}$,  
implies that each of those $\SF{n}_i$ must be of different values from $\SF{n}_k$, and thus 
cannot be represented by a finite number of distinct labeled nulls.

\subsection{Triangular-components of TGD extensions}

In contrast to $\tgds_2$ from Example \ref{cyc_aag_examp},
what we aim to achieve now is to identify syntactic conditions on 
TGDs so that such a ``distinguishable relation" of labeled nulls is limited in the chase derivation. 
As a consequence,
we end up with some nulls that need not be distinguishable from another, and as such, we can actually
{\em re-use} these nulls without introducing new ones in the chase derivation. This leads BCQ-Ans to be decidable
since TGDs in this class will have the FC property.

\begin{definition}[\textbf{\textit{TGD extension}}]\label{tgd_extension}
	Given a set of TGDs $\tgds$, we denote by $\tgds^+$ as the \emph{extension} of $\tgds$, 
	and is inductively defined as follows:
	
	\vspace*{-0.4cm}

	\begin{align}
		&\hspace*{-1.9cm}\tgds^0=\big\{\,\big\LA\BD(\tgd),\atomh\big\RA\,\mid\,
		\tgd\in\tgds\mbox{ and }\atomh\in\HD(\tgd)\,\big\};\label{tgds_ext_def_base}
	\end{align}

	\vspace*{-0.6cm}	

	\begin{align}
		&\hspace*{-3.6cm}\tgds^{i+1}=\,\tgds^i\,\cup\hspace*{3cm}\label{tgds_ext_def_ind_1}
    \end{align}	

	\vspace*{-0.6cm}			
		
	\begin{align}		
		&\hspace*{-0.1cm}\big\{\,\big\LA B_1\eta_1\cup B^*,\atomh_2\eta_2\big\RA\mbox{ \large $\mid$ }\LA B_1,\atomh_1\RA\in\tgds^0,\,
		       \LA B_2,\atomh_2\RA\in\tgds^i,\hspace*{-0.1cm}\label{tgds_ext_def_ind_1.5}\\
	    &\hspace*{0.2cm}\mbox{where }
		\eta_1:\VAR\big(B_1\cup\{\atomh_1\}\big)\longrightarrow\TERMS\big(B_1\cup\{\atomh_1\}\big)		
		\nonumber\\		
		&\hspace*{0.2cm}\mbox{and }\eta_2:\VAR\big(B_2\cup\{\atomh_2\}\big)\longrightarrow\TERMS\big(B_1\eta_1\cup B_2\cup\{\atomh_2\}\big)\nonumber\\
		&\hspace*{0.2cm}\mbox{such that:}\nonumber\\
		%
		%
		%
		&\hspace*{0.2cm}\mbox{(1) }\exists\atomb\in B_2\mbox{ such that }\atomh_1\eta_1=\atomb\eta_2
		 \mbox{ corresponds to}\label{tgds_ext_def_ind_1.5.5}\\
		&\hspace*{0.7cm}\mbox{the MGU of }\atomh_1\mbox{ and }\atomb,\mbox{ and }
		 \eta_1(X)=X,\mbox{ for each}\nonumber\\			
		&\hspace*{0.7cm}X\in\VAR(B_1)\setminus\VAR(\atomh_1);\nonumber\\		
		&\hspace*{0.2cm}\mbox{(2) }B^*= B_2\eta_2\setminus\{\atomb\eta_2\}\,\big\}.\label{tgds_ext_def_ind_2}
 \end{align}

	\vspace*{-0.2cm}

\end{definition}
\noindent
In order to avoid undesirable clashes in variables, we assume that each $\LA B_1,\atomh_1\RA\in\tgds^0$ and $\LA B_2,\atomh_2\RA\in\tgds^i$ mentioned in (\ref{tgds_ext_def_ind_1.5}) have the variables renamed so that $\VAR\big(B_1\cup\{\atomh_1\}\big)$ $\cap$ $\VAR\big(B_2\cup\{\atomh_2\}\big)$ $=$ $\emptyset$.
Then we set $\tgds^+$ $=$ $\tgds^\infty$ as the fixpoint of $\tgds^i$. We note that even though
$\tgds^+$ can be infinite in general, it follows from Theorem \ref{finite_steps_for_RTC} that it
is enough to consider just a finite number of iterations $\tgds^i$ to determine ``recursive triangular-components"
(as will be defined exactly in Definition \ref{triangular_comp_def}).

\comment
{
To keep the computation of $\tgds^+$ finite, we further assume that for each
$\LA B_1,H_1\RA$, $\LA B_2,H_2\RA$ $\in$ $\tgds^+$, $\LA B_1,H_1\RA$ ``homomorphic" to $\LA B_2,H_2\RA$
implies that $\LA B_1,H_1\RA$ $=$ $\LA B_2,H_2\RA$. Here, we say that  
$\LA B_1,H_1\RA$ \emph{homomorphic} $\LA B_2,H_2\RA$ $\in$ $\tgds^+$ if there exists a 
function $\theta:$ $\VAR(B_1\cup H_1)$ $\longrightarrow$ $\VAR(B_2\cup H_2)$ 
such that: 
(1) $\LA B_1\theta,H_1\theta\RA$ $=$ $\LA B_2,H_2\RA$; and 
(2) $r(t_1,\ldots,t_k)$ $\in$ $B_1$ \big(resp. $r(t_1,\ldots,t_k)$ $\in$ $H_1$\big) iff
$r(\theta(t_1),\ldots,\theta(t_k))$ $\in$ $B_2$ \big(resp. $r(\theta(t_1),\ldots,\theta(t_k))$ $\in$ $H_2$\big).
}

The TGD extension $\tgds^+$ of $\tgds$ contains as members pairs of sets of atoms and head atoms of the form 
``$\LA B,\atomh\RA$", respectively. Loosely speaking, the set $B$ represents the possible union of body atom of some 
TGD in $\tgds$ while $\atomh$ the head atom of some TGD that can be linked (transitively) through the repeated 
applications of the steps in (\ref{tgds_ext_def_ind_1})-(\ref{tgds_ext_def_ind_2}) (which is done until a fixpoint is reached).
The base case $\tgds^0$ in (\ref{tgds_ext_def_base}) first considers the 
pairs $\LA B,\atomh\RA$, where $B$ $=$ $\BD(\tgd)$ and $\atomh$ $\in$ $H$, for each $\tgd$ $\in$ $\tgds$.
Inductively, assuming we have already computed $\tgds^{i}$, we have that $\tgds^{i+1}$ is obtained by
adding the previous step $\tgds^{i}$ as well as adding the set 
as defined through (\ref{tgds_ext_def_ind_1})-(\ref{tgds_ext_def_ind_2}).

More specifically, using similar ideas to the \emph{TGD expansion} in \cite{CaliGP12} that was used in 
identifying the \emph{sticky-join} class of TGDs and \emph{tame reachability} in \cite{GottlobMP13}
used for the \emph{tame} class, 
the set (\ref{tgds_ext_def_ind_1})-(\ref{tgds_ext_def_ind_2}) considers the other head types
that can be (transitively) reached from some originating TGD. Indeed, as described in (\ref{tgds_ext_def_ind_1})-(\ref{tgds_ext_def_ind_2}), for $\LA B_1,\atomh_1\RA$ $\in$ $\tgds^0$ and 
$\LA B_2,\atomh_2\RA$ $\in$ $\tgds^{i}$ \big(i.e., as in (\ref{tgds_ext_def_ind_1.5})\big), 
we add the pair $\LA B_1\eta_1\cup B^*,\atomh_2\eta_2\RA$ into 
$\tgds^{i+1}$. Intuitively, with the homomorphism ``$\eta_2$" as described in 
(\ref{tgds_ext_def_ind_1})-(\ref{tgds_ext_def_ind_2}), the aforementioned pair $\LA B_1\eta_1\cup B^*,\atomh_2\eta_2\RA$
encodes the possibility that ``$\atomh_2\eta_2$" can be derived transitively from the union of bodies $B_1\eta_1$ and 
$B^*$ $=$ $B_2\eta_2\setminus \atomb\eta_2$. Here, we consider the \emph{union} of $B_1\eta_1$ and 
$B^*$ $=$ $B_2\eta_2\setminus \atomb\eta_2$ (instead of just $B_1\eta_1$) because it may take a combination of these atoms 
to reveal a possible ``recursive triangular-component" \footnote{We will define the notion of 
recursive triangular-components (RTC) precisely in Definition \ref{triangular_comp_def} of the following section.}.
We note that because $\eta_1$ only maps variables from $\VAR\big(B_1\cup\{\atomh_1\}\big)$ onto
$\TERMS\big(B_1\cup\{\atomh_1\}\big)$ for the condition $\atomh_1\eta_1$ $=$ $\atomb\eta_2$ 
\big(i.e., as in (\ref{tgds_ext_def_ind_1.5.5})\big), 
then we can track some of the originating variables from $B_1$ all the way through the head
``$\atomh_2\eta_2$", and which can be retained through iterative applications of the criterion given
in (\ref{tgds_ext_def_ind_1})-(\ref{tgds_ext_def_ind_2}). Importantly, we note that the connection
between $\atomh_1$ and $\atomb_2$ is inferred with $\atomh_1\eta_1$ $=$ $\atomb\eta_2$ 
(where $\atomb$ $\in$ $B_2\theta$) corresponding to the 
\emph{most general unifier} (MGU) of $\atomh_1$ and $\atomb$ (please see Condition (2) of set (\ref{tgds_ext_def_ind_1})-(\ref{tgds_ext_def_ind_2})).

\begin{example}\label{tgd_extension_examp}
	Let $\tgds_3$ be the following set of TGD rules:

	\vspace*{-0.5cm}		    

	\begin{align}
		\tgd_{31}:&\,\,\,\SF{t}(X,Y)\ra\exists Z\,\SF{t}(Y,Z),\nonumber\\
		\tgd_{32}:&\,\,\,\SF{t}(X,Y)\ra\,\SF{s}(X)\wedge\SF{s}(Y),\nonumber\\				
		\tgd_{33}:&\,\,\,\SF{t}(X_1,V)\wedge\SF{s}(V)\wedge\SF{t}(W,Z_1)\ra\SF{u}(X_1,V,W,Z_1),\nonumber\\
		\tgd_{34}:&\,\,\,\SF{u}(X_2,Y,Y,Z_2)\ra\SF{v}(X_2,Z_2),\nonumber\\
		\tgd_{35}:&\,\,\,\SF{v}(X_3,Z_3)\ra\SF{t}(X_3,Z_3).\nonumber							
	\end{align}

	\vspace*{-0.1cm}	
	
	\noindent
	Then from the rules $\tgd_{33}$, $\tgd_{34}$ and $\tgd_{35}$, we get the three pairs
	$p_{01}$ $=$ $\LA\{\SF{t}(X_1,V)$,$\SF{s}(V)$,$\SF{t}(W,Z_1)\}$,$\SF{u}(X_1,V,W,Z_1)\RA$,
	$p_{02}$ $=$ $\LA\{\SF{u}(X_2,Y,Y,Z_2)\}$,$\SF{v}(X_2,Z_2)\RA$ and 
	$p_{03}$ $=$ $\LA\{\SF{v}(X_3,Z_3)\}$, $\SF{t}(X_3,Z_3)\RA$		
	in $\tgds^0_3$, respectively. Then through the unification of head atom ``$\SF{v}(X_2,Z_2)$"
	of the pair $p_{02}$ and the body atom ``$\SF{v}(X_3,Z_3)$" of the pair $p_{03}$, then we 
	get the pair $p_{11}$ $=$ $\LA\{\SF{u}(X_2,Y,Y,Z_2),\SF{t}(X_2,Z_2)\}\RA$ in $\tgds^1_3$.
	Then finally, through the unification of the head atom ``$\SF{u}(X_1,V,W,Z_1)$" of 
	the pair $p_{01}$ and the body atom ``$\SF{u}(X_2,Y,Y,Z_2)$" of the pair $p_{11}$,
	then we further get the pair $p_{21}$ $=$ $\LA\{\SF{t}(X_1,V)$, $\SF{s}(V)$, $\SF{t}(V,Z_1)\}$, 
	$\SF{t}(X_1,Z_1)\RA$ $\in$ $\tgds^2_3$.
	%
	\comment
	{
	Then through the unification of head atom
	``$\SF{u}(X_1,V,W,Z_1)$" of rule $\tgd_{33}$ with the body atom ``$\SF{u}(X_2,Y,Y,Z_2)$" of $\tgd_{34}$, 
	we get the pair $p_{11}$ $=$ $\LA\{\SF{t}(X_1,V)$, $\SF{s}(V)$,
	$\SF{t}(V,Z_1)\}$, $\SF{v}(X_1,Z_1)\RA$ $\in$ $\tgds^1_3$.
	Then finally, through the unification of the head atom ``$\SF{v}(X_2,Z_2)$" of $\tgd_{34}$ and the
	body atom ``$\SF{v}(X_3,Z_3)$" of $\tgd_{35}$, we further get the pair $p_{21}$ $=$ 
	$\LA\{\SF{t}(X_1,V)$, $\SF{s}(V)$, $\SF{t}(V,Z_1)\}$, $\SF{t}(X_1,Z_1)\RA$ $\in$ $\tgds^2_3$.
	}
\end{example}
As will be seen in Definition \ref{triangular_comp_def}, the last pair $p_{21}$ in Example 
\ref{tgd_extension_examp} corresponds to what we will call a ``recursive triangular-component"
that will be defined precisely in Definition \ref{triangular_comp_def} in the following section.

\subsection{Triangularly-guarded TGDs}\label{triangularly-guarded_TGDs}

In this section, we introduce the key notion of triangularly-guarded TGDs,
which are so-called \emph{triangular-components}. 

We first introduce the notion of \emph{cyclically-affected} 
only variables in the body (i.e., set $B$) of some pair $\LA B,\atomh\RA$ $\in$ $\tgds^+$ where
$\tgds$ is a set of TGDs.
So towards this purpose, for a given pair $\LA B,\atomh\RA$ $\in$ $\tgds^+$, we define
$\widehat{\VAR}(\tgds,B)$ (i.e., ``$\widehat{\VAR}$" is read \emph{var-hat})
as the set of variables: 
$\big\{X$ $\mid$ $X$ $\in$ $\VAR(B)$
and $\bigcap\NULLSET(X,\tgd,\tgds)[B]$ $\cap$ $\CYCNULL(\tgds)$ $\neq$ $\emptyset\big\}$,
where $\bigcap\NULLSET(X,\tgd,\tgds)[B]$ denotes the intersection of the unions 
$\bigcup_{b\in\ARG(\atomb'),\,\atomb'\in\HD(\tgd'),\,\tgd'\in\tgds}\NULLSET(b,\atomb',\tgds)$, 
for each pair $(b,\atomb)$ such that $b$ $\in$ $\ARG(\atomb)\REST_X$ and $\atomb$ $\in$ $B$.
For convenience and when clear from the context, we simply refer to $\widehat{\VAR}(\tgds,B)$
as $\widehat{\VAR}(B)$.
Intuitively, variables in $\widehat{\VAR}(B)$ are placements for which an infinite 
number of labeled nulls can possibly be propagated in the set of atoms $B$ with respect to $\tgds$.
Loosely speaking, if $B$ $=$ $\BD(\tgd)$, for some set of TGDs $\tgds$, then $\widehat{\VAR}(B)$
contains the \emph{glut-variables} in \cite{KrotzschR11} that also fails the 
\emph{shyness} property \cite{LeoneMTV12} (please see Section 2 of this paper).

Next, we introduce the link variables
between ``body atoms''. Given two atoms $\atomb_1$, $\atomb_2$ $\in$ $B$, 
for some $\LA B,\atomh\RA$ $\in$ $\tgds^+$ where $\tgds$ is a set of TGDs, 
we set $\LINK(\tgds^+,B,\atomb_1,\atomb_2)$
\big(or just $\LINK(B,\atomb_1,\atomb_2)$ when clear from the context\big) as the set of variables
in the intersections 
$\big(\VAR(\atomb_1)$ $\cap$ $\VAR(\atomb_2)\big)$ $\cap$ $\widehat{\VAR}(\tgds,B)$.
Intuitively, $\LINK(B,\atomb_1,\atomb_2)$ denotes the cyclically-affected only variables
of $B$ that can actually ``join" (link) two common nulls between the body atoms $\atomb_1$ and $\atomb_2$ 
that can be obtained through some firing substitution.

Lastly, we now introduce the notion of variable markup.
Let $\atoma$, $\atomc$ and $\atoma'$ be three atoms such that $\REL(\atoma)$ $=$ $\REL(\atoma')$. 	 
Then similarly to \cite{CaliGP12}, we define the ``markup procedure" as follows. 
For the base case, we let $\atoma^0$ (resp. $\atomc^0$) denote the atom obtained from 
$\atoma$ (resp. $\atomc$) by marking each variable 
$X$ $\in$ $\VAR(\atoma)$ (resp. $X$ $\in$ $\VAR(\atomc)$) such that
$X$ $\notin$ $\VAR(\atomc)$ (resp. $X$ $\notin$ $\VAR(\atoma')$). 

Inductively, we define $\atoma^{i+1}$ (resp. $\atomc^{i+1}$) to be the atom 
obtained from $\atoma^{i}$ (resp. $\atomc^{i}$) 
as follows: for each variable $X$ $\in$ $\VAR(\atomc)$ (resp. $X$ $\in$ $\VAR(\atoma')$), 
if each variables in positions
$\ARG(\atomc)\REST_{X}$ (resp. $\ARG(\atoma')\REST_{X}$) occurs as marked in 
$\atomc^{i}$ (resp. $\atoma^{i}$), then each occurrence of $X$ is marked in 
$\atoma^{i}$ (resp. $\atomc^{i}$) to obtain the new atom $\atomc^{i+1}$
(resp. $\atoma^{i+1}$).
Then naturally, we denote by $\atoma^\infty$ (resp. $\atomc^\infty$) 
as the fixpoint of the markup applications. Finally, we denote by 
$\MVAR(\atoma,\atomc,\atoma')$ as the set 
of all the marked variables mentioned \emph{only} in $\atoma^\infty$
under atoms $\atomc$ and $\atoma'$ as obtained through the method above.

Loosely speaking, in the aforementioned variable markup, we can think of 
$\atoma$ as corresponding to some ``body atom" while $\atomc$ and 
$\atoma'$ as ``head atoms" that are reachable through the TGD extension 
$\tgds^+$ (see Definition \ref{tgd_extension}). 
Intuitively, the marked variables
represent element positions that may fail the sticky-join property, i.e., disappear in the recursion of the rules.
Intuitively, the sticky-join property insures decidability because only a finite number of elements can circulate
among the recursive application of the rules. 
As will be revealed in Definition \ref{triangular_comp_def},
we further note that we only consider marked variables in terms of the
triple $\LA\atoma,\atomc,\atoma'\RA$ because we only consider them for ``recursive triangular-components."

\begin{definition}[\textbf{\textit{Recursive triangular-components}}]\label{triangular_comp_def}
	Let $\tgds$ be a set of TGDs and $\tgds^+$ its extension as defined in Definition \ref{tgd_extension}.
	Then a \emph{recursive triangular-component} (RTC) ${\cal T}$ is a tuple
	
	\vspace*{-0.4cm}
	
	\begin{align}
		&\hspace*{-0.1cm}\big(\LA B,\atomh\RA,\LA\atoma,\atomb,\atomc\RA,\LA X,Z\RA,\atoma'\big),\label{triangular-component}
	\end{align}
	
	\vspace*{-0.1cm}	
	
	\noindent
	where: 
	\textbf{1.)} $\LA B,\atomh\RA$ $\in$ $\tgds^+$;
	\textbf{2.)} $\{\atoma,\atomb\}$ $\subseteq$ $B$, $\atoma$ $\neq$ $\atomb$ and
	             $\atomc$ $=$ $\atomh$;
	\textbf{3.)} $\atoma'$ is an atom and there exists a homomorphism $\theta:$ $\VAR(\atoma)$ $\longrightarrow$ $\VAR(\atoma')$
	             such that $\atoma\theta$ $=$ $\atoma'$ and either one of the following holds:
	\begin{description}
	    \item[\rm (a)] $\atomc$ $=$ $\atoma'$, or
	    \item[\rm (b)] there exists $\LA B',\atomh'\RA$ $\in$ $\tgds^+$ and function 
	             $\eta:$ $\VAR\big(B'\theta'\cup \{\atomh'\theta'\}\big)$ $\longrightarrow$ $\Gamma_{\cal C}\cup\Gamma_{\cal V}$,
	             where $\theta'$ is just a renaming substitution such that 
	             $\VAR\big(B'\theta'\cup\{\atomh'\theta'\}\big)$ $\cap$ $\VAR\big(B\cup\{\atomh\}\big)$ $=$ $\emptyset$, and where
	             $\atomc$ $\in$ $B'(\eta\circ\theta')$ and $\atoma'$ $=$ $\atomh'(\eta\circ\theta')$;   
	\end{description}	
	\textbf{4.)} $X$ and $Z$ are two distinct variables where $\{X,Z\}$ $\subseteq$ $\widehat{\VAR}(B)$, and
	             $X$ $\in$ $\VAR(\atoma)$, $Z$ $\in$ $\VAR(\atomb)$, $\{X,Z\}$ $\subseteq$ $\VAR(\atomc)$ 
	             and $X$ $\in$ $\VAR(\atoma')$; and lastly, 
	\textbf{5.)} there exists a tuple of distinct atoms 
	             $\vatomd$ $=$ $\atomd_{1}\ldots\atomd_{m}$ $\subseteq$ $B$ such that:
    \begin{description}
	    \item[\rm (a)] $\atoma$ $=$ $\atomd_{1}$ and $\atomd_{m}$ $=$ $\atomb$, and for each $i$ $\in$ $\{1,\ldots,m-1\}$,
	                   there exists $Y_{i}$ $\in$ $\LINK(B,\,\atomd_{i},\,\atomd_{i+1})$;	     
	    %
	    \item[\rm (b)] for some $i$ $\in$ $\{1,\ldots,m-1\}$, there exists
					    $Y'$ $\in$ $\LINK(B,\,\atomd_{i},\,\atomd_{i+1})\mbox{\large $\setminus$}\{X,Z\}$ 
					    such that $Y'$ $\in$ $\VAR(\atoma')$ implies 
					    all occurrences of variables in positions $\ARG(\atoma')\REST_{Y'}$
					    in the atom $\atoma$ are in $\MVAR\big(\atoma,\vatomc,\atoma')$.
    \end{description}
\end{definition}

\begin{figure}
	\centering 
	\hspace*{-0.5cm}
	\begin{tikzpicture}[scale=0.75, every node/.style={scale=0.6}]
	   \begin{scope}[shift={(3cm,-5cm)}, fill opacity=1]
	   	   \node at (-5.64,0.85){\LARGE $\atomd_{1}\,=\,\atoma$};	  
	   	   \node at (-6.6,0.62){\Large $Y_{1}\,\{$};
	   	   \node at (-6.08,0.35){\LARGE $\atomd_{2}$};	  
	   	   \node at (-6.2,0.08){\LARGE $\vdots$};
	   	   \node at (-5.85,-0.45){\LARGE $\atomd_{m-1}$};	   	   
	   	   \node at (-6.81,-0.66){\Large $Y_{m-1}\,\{$};	   	   
	   	   \node at (-5.65,-0.95){\LARGE $\atomd_{m}=\atomb$};  
	   	   \draw[thick,->] (-4.75,0.90) to (-2.5,0.2);
	   	   \draw[thick,->] (-4.65,-0.95) to (-2.5,-0.2);
	   	   \node at (-2.17,0.0){\LARGE $\tgds^+$};
	   	   \draw[thick,dotted,->] (-1.9,0) to (-1.5,0.0);	   	   
	   	   \draw[thick,->] (-1.4,0.0) to (-0.3,0.0);
	   	   \draw[thick,dotted,->] (0.55,0.0) to (2.08,0.0);	   	   	
	   	   \node[rotate=0] at (1.3,0.28){\Large $\LA B',\atomh'\RA$};	   	   
	   	   \draw[thick,dotted,->] (2.2,0.3) to [bend right=60] (-4.86,1.05);(1.7,0.05)
	   	   \node[rotate=-15] at (-3.3,0.85){\Large $\atoma\in B$};
	   	   \node[rotate=18] at (-3.2,-0.9){\Large $\atomb\in B$};	 
	   	   \node[rotate=0] at (-0.8,0.3){\Large $\atomc= \atomh$};
	   	   \node[rotate=-90] at (0.25,0.0){\Large $\{X,Z\}$};
	   	   \node[rotate=0] at (-0.1,0.0){\LARGE $\atomc$};
	   	   \node[rotate=-90] at (2.7,-0.03){\Large $X$};	   	   
	   	   \node[rotate=0] at (2.3,0.05){\LARGE $\atoma'$};	   	   	   	   	   	   
	   	   \node[rotate=0] at (-5.1,1.35){\Large $X$};	 
	   	   \node[rotate=0] at (-5.15,-1.4){\Large $Z$};	
	   	   \node[rotate=5] at (-2.1,2.75){\Large $\atoma\theta=\atoma'$};	   	   
	   	   \node[rotate=3] at (-2.1,2.2){\Large cycle};	   	   
	   \end{scope}
	\end{tikzpicture} 

	\vspace*{-0.3cm}
	
	\caption{ Recursive triangular-component (RTC).} 
	
	\vspace*{-0.1cm}
	        
    \label{triangular_comp_diagram} 
\end{figure}
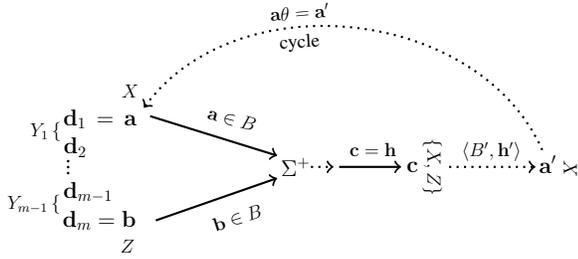

%
Generally, a recursive triangular-component (RTC) ${\cal T}$ of the form (\ref{triangular-component})
(see Definition \ref{triangular_comp_def} and Figure \ref{triangular_comp_diagram}), 
can possibly enforce an infinite cycle of labeled nulls being ``pulled" together into a relation in
the chase derivation. We explain this by using again the TGDs $\tgds_2$ $=$ $\{\tgd_{11},\tgd^*_{12}\}$
and database $D_2$ of Example \ref{cyc_aag_examp}. Here, let us assume that $B$ $=$ $\BD(\tgd^*_{12})$
and $\atomh$ $\in$ $\HD(\tgd^*_{12})$ such that $\LA B,\atomh\RA$ is the pair mentioned in (\ref{triangular-component}).
Then with the body atoms $\SF{t}(X,Y)$, $\SF{u}(Y,Z)$ $\in$ $\BD(\tgd^*_{12})$ 
and head atom $\SF{t}(X,Z)$ $=$ $\atomh$ $\in$ $\HD(\tgd^*_{12})$  
also standing for the atoms $\atoma$, $\atomb$ and $\atomc$ $=$ $\atomh$ in (\ref{triangular-component}), respectively, 
then we can form the RTC ${\cal T}$:

\vspace*{-0.5cm}

\begin{align}
	&\big(\big\LA B,\atomh\RA,\big\LA\SF{t}(X,Y),\SF{u}(Y,Z),\SF{t}(X,Z)\big\RA,\LA X,Z\RA,\SF{t}(X,Z)\big).
	\nonumber
\end{align}

\vspace*{-0.1cm}

We note here from Condition \textbf{3.)} of Definition \ref{triangular_comp_def} that the atom $\atomc'$
in (\ref{triangular-component}) is also the head atom ``$\SF{t}(X,Z)$", i.e., the choice (a) 
$\atomc$ $=$ $\atoma'$ of Condition \textbf{3.)} holds in this case.
For simplicity, we note that out example RTC ${\cal T}$ above retains the names of the variables ``$X$" and ``$Y$"
mentioned in (\ref{triangular-component}).
Loosely speaking, for two atoms $\SF{t}(\SF{n}_i,\SF{n}_j)$, $\SF{t}(\SF{n}_j,\SF{n}_k)$ 
$\in$ $\CHASE(D_2,\tgds_2)$, we have that rule $\tgd^*_{12}$ and its head atom ``$\SF{t}(X,Z)$"
would combine the two nulls ``$\SF{n}_i$" and ``$\SF{n}_j$" into a relation 
``$\SF{t}(\SF{n}_i,\SF{n}_j)$" in $\CHASE(D_2,\tgds_2)$. Since the variable ``$X$" is retained
in each RTC cycle via Condition \textbf{4.)} (see Figure \ref{triangular_comp_diagram}), this makes possible that
nulls held by ``$X$" in each cycle (in some substitution) to be pulled together into some other nulls held by ``$Z$"
as derived through the head atom ``$\SF{t}(X,Z)$". 

We further note that the connecting variable ``$Y$"
between the two body atoms ``$\SF{t}(X,Y)$" and ``$\SF{u}(Y,Z)$" corresponds to the variables $Y_i$
$\in$ $\LINK(B,\,\atomd_{i},\,\atomd_{i+1})$ of point (a) of Condition \textbf{5.)}, 
and for some $i$, some $Y'$ $\in$ $\LINK(B,\,\atomd_{i},\,\atomd_{i+1})\mbox{\large $\setminus$}\{X,Z\}$  
also appears as marked (i.e., $Y'$ $\in$ $\MVAR\big(\atoma,\atomc,\atoma')$) in 
point (b) of Condition \textbf{5.)} with respect to the atom $\atoma'$. 
Intuitively, we require in (b) of Condition \textbf{5.)} that \emph{some} of these variables $Y'$ occur as marked 
(w.r.t. $\atoma'$) so that labeled nulls of some link variables have a chance to disappear in the RTC cycle,
otherwise, they can only link and combine a bounded number of labeled nulls due to the 
\emph{sticky-join} property \cite{CaliGP12}.

\begin{example}\label{RTC_examp}
	Consider again the pair $p_{21}$ $=$ $\LA\{\SF{t}(X_1,V)$, $\SF{s}(V)$, $\SF{t}(V,Z_1)\}$, $\SF{t}(X_1,Z_1)\RA$ $\in$ $\tgds^2_3$ from Example \ref{tgd_extension_examp}. 
	Then with the pair $p_{21}$ standing for $\LA B,\atomh\RA$ in (\ref{triangular-component}),
	the atoms ``$\SF{t}(X_1,V)$", ``$\SF{t}(V,Z_1)$", ``$\SF{t}(X_1,Z_1)$" and ``$\SF{t}(X_1,Z_1)$" for the atoms 
	$\atoma$, $\atomb$, $\atomc$ and $\atomc'$ in (\ref{triangular-component}), respectively, and
	variables $\LA X_1,Z_1\RA$ for the variables $\LA X,Z\RA$ in (\ref{triangular-component}), then we can get
	a corresponding RTC ${\cal T}_1$ $=$ 
	$\big(p_{21},\big\LA \SF{t}(X_1,V),\SF{t}(V,Z_1),\SF{t}(X_1,Z_1)\big\RA,\LA X_1,Z_1\RA,\SF{t}(X_1,Z_1)\big)$
	as illustrated in Figure \ref{triangular_comp_examp_diagram}.	
\end{example}
\begin{figure}
	\centering 
	\begin{tikzpicture}[scale=0.75, every node/.style={scale=0.6}]
	\begin{scope}[shift={(3cm,-5cm)}, fill opacity=1]
	\node at (-5.64,0.85){\Large $\SF{t}(X_1,V)$};	  
	\node at (-6.78,0.5){\Large $V$};
	\node at (-5.85,0.0){\Large $\SF{s}(V)$};
	\draw[decorate,decoration={brace,mirror,raise=7pt}] (-6,0.9) to (-6,0.1);
	\draw[decorate,decoration={brace,raise=7pt}] (-6,-0.9) to (-6,0.0);		   	   
	\node at (-6.78,-0.5){\Large $V$};	   	   
	\node at (-5.65,-0.95){\Large $\SF{t}(V,Z_1)$};	   
	\draw[thick,->] (-4.95,0.90) to (-2.5,0.2);
	\draw[thick,->] (-4.95,-0.95) to (-2.5,-0.2);
	\node at (-2.17,0.0){\LARGE $\tgds^+_3$};
	\draw[thick,dotted,->] (-1.9,0) to (-1.5,0.0);	   	   
	\draw[thick,->] (-1.4,0.0) to (1.0,0.0);
	\draw[thick,dotted,->] (1.1,0.45) to [bend right=60] (-4.9,1.15);
	\node[rotate=-20] at (-3.4,0.8){\large $\SF{t}(X_1,V)\in B$};
	\node[rotate=18] at (-3.4,-0.9){\large $\SF{t}(V,Z_1)\in B$};	 
	\node[rotate=0] at (-0.2,0.25){\large $\SF{t}(X_1,Z_1)= \atomh$};		   	   
	\node[rotate=-90] at (2.65,0.0){\Large $\{X_1,Z_1\}$};
	\node[rotate=0] at (1.75,0.05){\Large $\SF{t}(X_1,Z_1)$};	   	   	   	   
	\node[rotate=0] at (-5.3,1.35){\Large $X_1$};	 
	\node[rotate=0] at (-5.3,-1.4){\Large $Z_1$};	
	\node[rotate=0] at (0.2,2.7){\Large $\SF{t}(X_1,V)\theta=\SF{t}(X_1,Z_1)$, 
		                          where $\theta$ $:=$ $\{X_1\mapsto X_1, V\mapsto Z_1\}$};	   	   
	\node[rotate=-3] at (-2.1,2.1){\Large cycle};	   	   
	\end{scope}
	\end{tikzpicture} 

	\vspace*{-0.2cm}
	
	\caption{RTC ${\cal T}_1$ of $\tgds_3$ with $\LA B,\atomh\RA$ $=$ $p_{21}$ of Example \ref{tgd_extension_examp}.} 
	
	\vspace*{-0.4cm}
	
	\label{triangular_comp_examp_diagram} 
\end{figure}

\begin{definition}[\textbf{\textit{Triangularly-guarded TGDs}}]\label{TG_def}
    A set of TGDs $\tgds$ is \emph{triangularly-guarded} (TG) iff for each
    RTC ${\cal T}$ of the form (\ref{triangular-component}) 
    (see Definition \ref{triangular_comp_def} and Figure \ref{triangular_comp_diagram}),
    we have that there exists some atom $\atomd$ $\in$ $B$ such that 
    $\{X,Z\}$ $\subseteq$ $\VAR(\atomd)$.
\end{definition}
\noindent
For convenience, we denote by $\TG$ as the class of all the triangularly-guarded TGDs.

\begin{example}
	Consider again the TGDs $\tgds_1$ in Example \ref{motive_examp} containing rules $\tgd_{11}$ and $\tgd_{12}$. 
	Then because there cannot be any pair $\LA B,\atomh\RA$ $\in$ $\tgds^+_1$ that would combine the two variables 
	``$X$" and ``$Z$" of rule $\tgd_{12}$ into a single head atom in $\tgds^+_1$, then it follows that $\tgds_1$ 
	cannot have any RTC. Therefore, it trivially follows from Definition \ref{TG_def} that $\tgds_1$ is in the class $\TG$.
\end{example}

\section{Main Results}

We now examine the important properties of the $\TG$ class of TGDs. 
In particular, we show that membership decision and BCQ-Ans under this new class $\TG$ of TGDs are both decidable, 
with the latter because $\TG$ satisfies the FC property.

As we mentioned within the paragraph right after Definition \ref{tgd_extension}, it is actually sufficient to only consider a finite
number of iterations of the TGD expansion $\tgds^i$ to determine for the existence of RTC (otherwise, it can be the case that
determining membership for the class $\TG$ is undecidable). 
So towards this purpose, for a set of TGDs $\tgds$ of schema ${\cal R}$,
we define the following number ``${\cal B}(\tgds)$" as follows:
%
%
%

\vspace*{-0.4cm}

\begin{align}
	& 
	 {\cal B}(\tgds)\,=\,\SF{maxb}^{\,[|\tgds^0|\cdot|{\cal R}|\cdot\SC{bell}(\SF{maxa}+\CONST(\tgds))]}
	\label{bell_bound}
\end{align}

\vspace*{-0.2cm}

\noindent
where: \footnote{Recall that ``$\tgds^0$" is the base case of $\tgds^+$ in Definition \ref{tgd_extension}.}
(1) $\SF{maxb}$ $=$ $\SC{max}\{\,|\BD(\tgd)|$ $\mbox{\Large $\mid $ }$ $\tgd\in\tgds\,\}$ 
   (i.e., the largest size of body atoms mentioned in a rule of $\tgds$);
(2) $\SF{maxa}$ $=$ $\SC{max}\{\,|r|$ $\mid$ $r\in{\cal R}\,\}$ (i.e., the maximum arity of a relation $r$ $\in$ $|{\cal R}|$); 
(3) $\TERMS(\tgds)$ is the set of terms mentioned in $\tgds$; and
(4) for a number $n$ $\in$ $\mathbb{N}$,
    $\SC{bell}(n)$ denotes the \emph{bell number} of a set of size $n$,
    (which is a number exponential to $n$). 
    More specifically, the bell number $\SC{bell}(n)$ can be defined
    through the \emph{recursive} definition $\SC{bell}(n+1)$ $=$ $\sum_{k=0}^n \binom nk\cdot\SC{bell}(k)$,
    where $\SC{bell}(0)$ $=$ $1$. 

Intuitively, what we aim to achieve with the number ${\cal B}(\tgds)$ is to set a bound on the iterations of
$\tgds^+$ (where we recall that $\tgds^+$ $=$ $\tgds^\infty$) so that we can be sure that there exists an
RTC ${\cal T}$ $\in$ $\tgds^+$ iff there also exists one in ${\cal T}'$ $\in$ $\tgds^{{\cal B}(\tgds)}$.
The main idea of why it is sufficient to consider the number ${\cal B}(\tgds)$ as the bound for the iterations
of $\tgds^i$ is related to the ideas of equivalent atom types in \cite{ChenLZZ11}. Thus, borrowing ideas
from \cite{ChenLZZ11}, given two atoms $\atoma_1$ and $\atoma_2$,
we say that $\atoma_1$ and $\atoma_2$ are \emph{type-equivalent}, denoted $\atoma_1$ $\sim$ $\atoma_2$,
iff there exists an isomorphism (i.e., bijective mapping) $h:$ $\TERMS(\atoma_1)$ $\longrightarrow$ $\TERMS(\atoma_2)$, 
where the following conditions are satisfied: 
(1) $\REL(\atoma_1)$ $=$ $\REL(\atoma_2)$;
(2) $h(a)$ $=$ $a$ $=$ $h^{-1}(a)$, for each $a$ $\in$ $\big(\CONST(\atoma_1)$ $\cup$ $\CONST(\atoma_2)\big)$; and 
(3) $h(X)$ $=$ $X$ $=$ $h^{-1}(X)$, for each $X$ $\in$ $\VAR(\atoma_1)$ $\cap$ $\VAR(\atoma_2)$. 

Intuitively,
$\atoma_1$ $\sim$ $\atoma_2$ means that $\atoma_1$ and $\atoma_2$ are isomorphic and agrees on the (argument) 
positions mentioning the constants and shared variables between $\atoma_1$ and $\atoma_2$. 
In relation to the number ${\cal B}(\tgds)$ described in (\ref{bell_bound}), the exponent 
``$|\tgds^0|$ $\cdot$ $[|{\cal R}|$ $\cdot$ $\SC{bell}(\SF{maxa}+\CONST(\tgds))]$" considers the size of all 
type-equivalent atoms since ``$\SC{bell}(\SF{maxa}+\CONST(\tgds))$" first considers the possible
type-equivalent atoms taking the constants into account, while factor ``$|\tgds^0|$ $\cdot$ $|{\cal R}|$" further considers
those atom types for each of the possible relation symbols and as mentioned per each rule of $\tgd$. 
Finally, we raise ``$\SF{maxb}$" to the power
of the exponent (i.e., ``$|\tgds^0|$ $\cdot$ $[|{\cal R}|$ $\cdot$ $\SC{bell}(\SF{maxa}+\CONST(\tgds))]$") since
we have to consider the derivation for each possible body atom in the rules.

%

\begin{theorem}[\textbf{\textit{Decidable class membership}}]\label{finite_steps_for_RTC}
	Let $\tgds$ be a set of TGDs and ${\cal B}(\tgds)$ the number as defined in (\ref{bell_bound}) above.
	Then there exists an RTC ${\cal T}$ $\in$ $\tgds^+$ (see Definition \ref{tgd_extension}) iff 
	there exists an RTC ${\cal T}'$	$\in$ $\tgds^{\,{\cal B}(\tgds)}$.
\end{theorem}
\begin{proofsketch} We prove the contraposition: 
	``\textit{there is no RTC ${\cal T}$ $\in$ $\tgds^+$ iff 
		there is no RTC ${\cal T}'$ $\in$ $\tgds^{\,{\cal B}(\tgds)}$.}"  

	(``$\Longrightarrow$") Clearly, because $\tgds^{\,{\cal B}(\tgds)}$ $\subseteq$ $\tgds^+$ $=$ $\tgds^\infty$
	(see Definition \ref{tgd_extension}), then it follows from Definition \ref{TG_def} that 
	there does not exists an RTC ${\cal T}$ $\in$ $\tgds^+$ implies there does not exists any RTC 
	${\cal T}'$ $\in$ $\tgds^{\,{\cal B}(\tgds)}$ as well.
	
	(``$\Longleftarrow$") Assume that there does not exists an RTC ${\cal T}'$ $\in$ $\tgds^{\,{\cal B}(\tgds)}$. 
	Now let us call a 
	\emph{path} of length $k$ in $\tgds^0$ (i.e., the initial step of $\tgds^\infty$), as a sequence of pairs:
	$\big(\LA B_1,\atomh_1\RA$, $\ldots$, $\LA B_k,\atomh_k\RA\big)$, where for each $i$ $\in$ $\{1$, $\ldots$, $k-1\}$,
	there exists some atom $\atomb_{i+1}$ $\in$ $B_{i+1}$ such that $\atomh_i$ and $\atomb_{i+1}$ are unifiable. Then
	we have that such a path actually corresponds to the derivation sequence of the sequence of steps in computing $\tgds^+$.
	Intuitively, at each step $\tgds^j$ (as described through Definition \ref{tgd_extension}), what we actually do is 
	``unfold" these atoms $\atomb_{i+1}$ $\in$ $B_{i+1}$ to reveal if they can possibly lead to an RTC. On the other hand,
	because the number ${\cal B}(\tgds)$ would consider path of lengths ${\cal B}(\tgds)$ in $\tgds^{{\cal B}(\tgds)}$, then
	we can be sure that we had considered (i.e., unfolded) all the possible atom types that will be encountered along the path
	up to type-equivalence, ``$\sim$. " Therefore, since $\tgds^{{\cal B}(\tgds)}$ does not have an RTC, then it follows that 
	$\tgds^+$ cannot have an RTC as well.	
\end{proofsketch}

\vspace*{-0.4cm}

\begin{theorem}[\textbf{\textit{Membership complexity}}]\label{membership_complexity}
	Deciding whether $\tgds$ $\in$ $\TG$ is in $2$-\SC{ExpTime} but is
	\SC{Pspace}-hard in general.		
\end{theorem}

\vspace*{-0.2cm}

\begin{proof} (``\textit{upper-bound}") Based on the number ${\cal B}(\tgds)$ as defined in 
	(\ref{bell_bound}) above, the upper-bound follows by only selecting the pairs 
	$\LA B,\atomh\RA$ $\in$ $\tgds^{i}$ at each step in (\ref{tgds_ext_def_ind_1.5})-(\ref{tgds_ext_def_ind_2})
	(see Definition \ref{tgd_extension}) for which their corresponding
	path has length $k$ $\leq$ ${\cal B}(\tgds)$ (see proof of Theorem \ref{finite_steps_for_RTC}
	for the notion of ``path"). Then through this restriction, we can get that $|\tgds^{{\cal B}(\tgds)}|$
	will be within order $O(2^{2^{|{\cal R}|\cdot|\tgds|\cdot\SF{maxa}\cdot\SF{maxb}\cdot\TERMS(\tgds)}})$. 
	%
	
	(``\textit{lower-bound}") Now we show \SC{Pspace}-hardness from ``first-principles" as follows.
	Let $L$ be an arbitrary decision problem in \SC{Pspace}. Then from the definition
	of complexity class \SC{Pspace} \cite{Papadimitriou:complexity}, 
	there exists some \emph{deterministic Turing machine}
	$M$ such that for any string $\vect{s}$, $\vect{s}$ $\in$ $L$ iff $M$ accepts
	$\vect{s}$ using at most $p(|\vect{s}|)$ steps for some polynomial $p(n)$.
	Consider a Turing machine $M$ to be the tuple
	$\LA Q$, $\Gamma$, $\Box$, $\Sigma$, $\delta$, $q_0$, $F\RA$, where
	(1) $Q$ $\neq$ $\emptyset$ is a finite set of states; 
	(2) $\Gamma$ $\neq$ $\emptyset$ is a finite set of alphabet symbols;
	(3) $\Box$ $\in$ $\Gamma$ is the ``blank" symbol; 
	(4) $\rhd$ $\in$ $\Gamma$ is the ``left-end-marker" symbol;
	(5) $\Sigma$ $\subseteq$ $\Gamma$ $\setminus$ $\{\Box,\,\rhd\}$ is the set of input symbols; 
	(6) $\delta$ $:$ $(Q\setminus F)$ $\times$ $\big(\Gamma\setminus\{\rhd\}\big)$ $\longrightarrow$ $Q$ $\times$
	$\Gamma$ $\times$ $\{L,R\}$ is the transition function; 
	(7) $q_0$ $\in$ $Q$ is the initial state; and lastly, 
	(8) $F$ $=$ $(F_{accept}$ $\cup$ $F_{reject})$ $\subseteq$ $Q$ is the set of final states such that
	$F_{accept}$ $\cap$ $F_{reject}$ $=$ $\emptyset$ and $F_{accept}$ ($F_{reject}$) is the accepting 
	(rejecting) states.	
	Then given an input string $\vect{s}$ $=$ $c_1\ldots c_l$
	we define the TGDs $\tgds_{S}(\vect{s})$, $\tgds_{R}(\vect{s})$, 
	$\tgds_{L}(\vect{s})$ and $\tgds_{A}(\vect{s})$ 
	as follows:

	\vspace*{-0.4cm}
	
	\begin{align}
		&\hspace{-4.1cm}\tgds_{S}(\vect{s})=\nonumber 
	\end{align}
	
	\vspace*{-1.0cm}
	
	\begin{align}			
	    &\big\{\SF{t}(X,Y)\wedge\SF{t}(Y,Z)\ra\SF{cf}(X,Y,Z,q_0,c_1,\ldots,c_l,\overbrace{\Box,\ldots,\Box}^{p(|\vect{s}|)-l},0)\big\};
		\nonumber
	\end{align}	

	\vspace*{-0.3cm}	
		
	\begin{align}		
		%
		%
		&\tgds_{R}(\vect{s})=\big\{\SF{cf}(X,Y,Z,q,T_1,\ldots,T_{k-1},a,T_{k+1},\ldots,T_{p(|\vect{s}|)},k)\nonumber\\
		&\hspace*{0.2cm}\ra\SF{cf}(X,Y,Z,q',T_1,\ldots,T_{k-1},b,T_{k+1},\ldots,T_{p(|\vect{s}|)},k+1)\big\}\nonumber 
	\end{align}	

	\vspace*{-0.2cm}	
		
	\begin{align}
		&\tgds_{L}(\vect{s})=\big\{\SF{cf}(X,Y,Z,q,T_1,\ldots,T_{k-1},a,T_{k+1},\ldots,T_{p(|\vect{s}|)},k)
	    \nonumber\\
		&\hspace*{0.2cm}\ra\SF{cf}(X,Y,Z,q',T_1,\ldots,T_{k-1},b,T_{k+1},\ldots,T_{p(|\vect{s}|)},k-1)\big\}\nonumber 
	\end{align}	
	\begin{align}
		&\tgds_{A}(\vect{s})=\big\{\SF{cf}(X,Y,Z,q,T_1,\ldots,T_{k-1},T_k,T_{k+1},\ldots,T_{p(|\vect{s}|)},k)\nonumber\\
		&\hspace*{1.1cm}\ra\SF{t}(X,Z)\mbox{\Large $\mid$ } q\in F\mbox{ and }1< k\leq p(|\vect{s}|)\big\}.
		\nonumber
	\end{align}	
	Here, we have that ``$\tgds_{S}(\vect{s})$" corresponds to the starting configuration, ``$\tgds_{R}(\vect{s})$"
	the right-transition movements $\delta(q,a)$ $=$ $(q',b,R)$,  
	``$\tgds_{L}(\vect{s})$" the left-transition movements $\delta(q,a)$ $=$ $(q',b,L)$ and 
	``$\tgds_{A}(\vect{s})$" the accepting states of the machine $M$ under the input string $\vect{s}$.
		
	Then with
	$\tgds$ $=$ $\big\{\SF{t}(X,Y)\ra\exists Z\,\SF{t}(Y,Z)\big\}$ $\cup$ $\tgds_{S}(\vect{s})\cup\tgds_{R}(\vect{s})$ $\cup$ $\tgds_{L}(\vect{s})\cup\tgds_{A}(\vect{s})$,	
	it follows that $M$ accepts $\vect{s}$ iff $\tgds$ $\notin$ $\TG$. We note here that we added the 
	TGD rule ``$\SF{t}(X,Y)\ra\exists Z\,\SF{t}(Y,Z)$" to $\tgds$ so that $\tgds$ is not 
	\emph{weak acyclic} (WA).	
\end{proof}


Now we show that the new $\TG$ class satisfies the finite-controllability (FC) property. So towards this purpose,
we first introduce the notion of interchangeable nulls, which will play the key role in proving the
FC property of the class $\TG$ of TGDs.

\begin{definition}[\textbf{\textit{Interchangeable nulls}}]\label{int_nulls}
	Let $\vatoma$ $=$ $\atoma_1\ldots\atoma_l$ be a tuple of atoms where $\TERMS(\vatoma)$ $\subseteq$ $\Gamma_{\cal V}$, 
	$D$ a database, $\tgds$ a set of TGDs 
	and $\SF{n}_{i}$, $\SF{n}_{j}$ $\in$ $\Gamma_{\SF{N}}$ \big($i,j$ $\in$ $\mathbb{N}$\big).
	Then we say that $\SF{n}_{i}$ and $\SF{n}_{j}$ are $\vatoma$-\emph{interchangeable} under 
	$\CHASE(D,\tgds)$ if for each \emph{connected} tuple of atoms 
	$\vatoma\theta$ $=$ $\theta(\atoma_1)\ldots\theta(\atoma_l)$ $\subseteq$ $\CHASE(D,\tgds)$, 
	where $\theta$ is a bijective (renaming) substitution,
	we have that $\{\SF{n}_{i},\SF{n}_{j}\}$ $\subseteq$ $\NULLS(\vatoma\theta)$
	implies 
	there exists a homomorphism $\theta':$ $\NULLS(\vatoma\theta)$ $\longrightarrow$ $\Gamma_\SF{N}$
	such that: (1) $\theta'(\SF{n}_{i})$ $=$ $\theta'(\SF{n}_{j})$; and 
	(2) $\vatoma(\theta'\circ\theta)$ $\subseteq$ $\CHASE(D,\tgds)$. 	
\end{definition}

Intuitively, with the tuple of atoms $\vatoma$ $=$ $\atoma_1\ldots\atoma_l$ as above, 
we have that $\SF{n}_i$ and $\SF{n}_j$ are ``$\vatoma$-interchangeable" 
under $\CHASE(D,\tgds)$ guarantees that if for some BCQ $Q$ $=$ $\exists\vect{X}\varphi(\vect{X})$ $\ra$ $q$ 
we have that $\CHASE(D,\tgds)$ $\models$ $Q$, then if $\varphi$ $=$ $\theta(\atoma_1)\wedge\ldots\wedge\theta(\atoma_l)$
for some renaming substitution $\theta$ (i.e., $\vatoma$ is the same ``type" as $\varphi$), 
then we have that simultaneously replacing all occurrences of $\SF{n}_j$ by $\SF{n}_i$ in $\CHASE(D,\tgds)$ 
would not affect the fact that $\CHASE(D,\tgds)$ $\models$ $Q$. 

Before we present the following Theorem \ref{bounded_nulls}, it is necessary to firstly introduce the notion \emph{level} in a chase that we define inductively as follows \cite{CaliGP12}: (1 ) for an atom $\atoma$ $\in$ $D$, we set $\LEVEL(\atoma)$ $=$ $0$; then inductively,
(2) for an atom $\atoma$ $\in$ $\CHASE(D,\tgds)$ obtained via some chase step $I_{k}\xrightarrow{\sigma,\,\eta}I_{k+1}$,
we set $\LEVEL(\atoma)$ $=$ $\MAX\big(\big\{\LEVEL(\atomb) \mid \atomb\in\BD(\tgd\eta)\}\big)$ $+$ $1$. Then finally,
for some given $k$ $\in$ $\mathbb{N}$, we set 
$\CHASE^k(D,\tgds)$ $=$ $\big\{\atoma\mid\atoma\in\CHASE(D,\tgds)\mbox{ and }\LEVEL(\atoma)\leq k\big\}$.
Intuitively, $\CHASE^k(D,\tgds)$ is the instance containing atoms that can be derived in a fewer or equal to 
$k$ chase steps. 

\begin{theorem}[\textbf{\textit{Bounded nulls}}]\label{bounded_nulls}
	Let $D$ be a database and $\tgds$ $\in$ $\TG$. Then for each
	tuple of atoms $\vatoma$, there exists numbers $N$, $N'$ $\in$ $\mathbb{N}$, where $N'$ $>$ $N$,  
	and such that $\forall k$ $\in$ $\mathbb{N}$, we have that 
	$\SF{n}_j$ $\in$ $\big[\NULLS\big(\CHASE^{\,N+N'+k}(D,\tgds)\big)$ $\mbox{\Large $\setminus$}$ $\NULLS\big(\CHASE^{\,N+N'}(D,\tgds)\big)\big]$ implies 
	$\exists\SF{n}_i$ $\in$ $\NULLS\big(\CHASE^{\,N}(D,\tgds)\big)$ where $\SF{n}_i$ and $\SF{n}_j$ 
	are $\vatoma$-interchangeable under $\CHASE(D,\tgds)$.	
\end{theorem}
\begin{proofsketch} 
	%
	As the full proof is quite technical and tedious, for space reasons, we will only provide a sketch that outlines
	the main ideas of the proof.
	
	Let us assume on the contrary that for some database $D$ and $\tgds$ $\in$ $\TG$, we have that
	the following condition holds:
	\begin{align}
		&\exists\vatoma,\,\forall N,N'\in\mathbb{N},\,\exists k\in\mathbb{N},\,\exists\SF{n}_j\in\Gamma_\SF{N},\,
		\forall\SF{n}_i\in\Gamma_\SF{N},\mbox{ where:}\nonumber\\ 
		&\SF{n}_j\in\big[\NULLS\big(\CHASE^{\,N+N'+k}(D,\tgds)\big)\mbox{\Large $\setminus$}\nonumber\\
		&\hspace*{1cm}\NULLS\big(\CHASE^{\,N+N'}(D,\tgds)\big)\big],
		 \nonumber\\
		&\SF{n}_i\in\NULLS\big(\CHASE^{\,N}(D,\tgds)\big)\mbox{ and }\SF{n}_i\mbox{ and }\SF{n}_j\mbox{ are \emph{not} }\nonumber\\
	    &\vatoma\mbox{-interchangeable under }\CHASE(D,\tgds).\label{contra_condition}
	\end{align}

	\vspace*{-0.2cm}	
	
	\noindent	 
	In particular, since this holds for all $N,N'$ $\in$ $\mathbb{N}$, then we can set a particular $N'$ to be the value:
	$N'$ $=$ $N^N$, where:

	\vspace*{-0.5cm}
	
	\begin{align}
		N\,=\,m^{(\,|\DOM(D)|+|\CONST(\tgds)|+|\VAR(\vatoma)|\,)},\label{N-val} 	
	\end{align}	

	\vspace*{-0.3cm}	
	
	\noindent	
	and where we further have the number $m$ to be:
	%
	
	\vspace*{-0.5cm}	
	
	\begin{align}
		m\,=\,d&\cdot\MAX\big\{\big|\VAR(\LA B,\atomh\RA)\big|\mid\LA B,\atomh\RA\in\tgds^{{\cal B}(\tgds)}\big\}
		      \nonumber\\
		&\cdot\MAX\big\{|r|\mid r\in{\cal R}\big\}\cdot\big|\VAR(\vatoma)\big|,\label{m-val}
	\end{align}	

	\vspace*{-0.2cm}	
	
	\noindent	
	where $d$ $=$ $|\tgds^0|\cdot|{\cal R}|\cdot\SC{bell}(\SF{maxa}+\CONST(\tgds))$
	(i.e., the exponent of the number ``${\cal B}(\tgds)$" in (\ref{bell_bound})).	
	Here, the value ``$m^{(\,|\DOM(D)|+|\CONST(\tgds)|+|\VAR(\vatoma)|\,)}$" considers the possible length of a 
	sequence of chase derivation steps under $\tgds^{{\cal B}(\tgds)}$ and w.r.t. to the size of the database $D$ as well
	as the tuple of atoms $\vatoma$. Indeed, the number 
	$d$ $\cdot$ $\MAX\big\{\big|\VAR(\LA B,\atomh\RA)\big|\mid\LA B,\atomh\RA\in\tgds^{{\cal B}(\tgds)}\big\}$ 
	$\cdot$ $\MAX\big\{|r|\mid r\in{\cal R}\big\}$ $\cdot$ $|\VAR(\vatoma)|$
	considers all the ``ground" instances of possible rules that can be applied, while the exponent
	``$|\DOM(D)|+|\CONST(\tgds)|+|\VAR(\vatoma)|$" (through which ``$m$" is raised by) considers all the 
	possible instantiations under $\DOM(D)$ $\cup$ $\CONST(\tgds)$. 
	We also factor in
	the size of $|\VAR(\vatoma)|$ as this will be needed in our arguments below. 
	Then given the condition
	in (\ref{contra_condition}), we can build an infinite sequence of distinct labeled nulls: 
	$\SF{n}_{k_{j_0}}$, $\SF{n}_{k_{j_1}}$, $\SF{n}_{k_{j_2}}$, $\SF{n}_{k_{j_3}}$, $\ldots$, 
	that satisfies the following conditions:
	\begin{enumerate}
        \item $k_{j_0}$ $<$ $k_{j_1}$ $<$ $k_{j_2}$ $<$ $k_{j_3}$, $\ldots$;\label{property_1}
		\item $N$ $=$ $j_0$ $<$ $j_1$ $<$ $j_2$ $<$ $j_3$, $\ldots$;\label{property_2}
		\item for each $i$ $\in$ $\{1,2,3,\ldots,\}$ $=$ $\mathbb{N}$, we have that:\\ 
		      $\SF{n}_{k_{j_0}}$ $\in$ $\NULLS\big(\CHASE^{\,j_{0}}(D,\tgds)\big)$ and
		      $\SF{n}_{k_{j_i}}$ $\in$ $\big[\NULLS\big(\CHASE^{\,j_i}(D,\tgds)\big)
		             \mbox{\Large$\setminus$}\NULLS\big(\CHASE^{\,j_{i-1}}(D,\tgds)\big)\big]$, 
		      and there exists a bijective substitution $\theta_i$ on $\vatoma$
		      where:\label{property_3} 
		      \begin{enumerate}
		      	\item $\vatoma\theta_i$ is $\NULLS$-connected;\label{property_3_1}
		      	\item $\vatoma\theta_i$ $\subseteq$ $\CHASE(D,\tgds)$;\label{property_3_2}
		      	\item $\{\SF{n}_{k_{j_0}},\SF{n}_{k_{j_i}}\}$ $\subseteq$ $\NULLS(\vatoma\theta_i)$;\label{property_3_3}
		      	\item there does not exists a homomorphism 
		      	      $\theta':$ $\NULLS(\vatoma\theta_i)$ $\longrightarrow$ $\Gamma_\SF{N}$ such that:
		      	      (1) $\theta'(\SF{n}_{k_{j_0}})$ $=$ $\theta'(\SF{n}_{k_{j_i}})$; and 
		      	      (2) $\vatoma(\theta'\circ\theta_i)$ $\subseteq$ $\CHASE(D,\tgds)$.
		      	\label{property_3_4}
		      \end{enumerate}
	\end{enumerate}
	Then given a large enough $\mathbf{i}$ $\in$ $\{1,2,3,\ldots\}$ $=$ $\mathbb{N}$ such that
	$\mathbf{i}$ $>$ $N'$ $=$ $N^N$ holds, (i.e., \emph{doubly-exponential} to $|D|$ and $|\tgds|$), 
	we can choose the corresponding bijective homomorphism $\theta_\mathbf{i}$, 
	where $\vatoma\theta_\mathbf{i}$ $\subseteq$ $\CHASE(D,\tgds)$ and
	$\{\SF{n}_{k_{j_0}},\SF{n}_{k_{j_\mathbf{i}}}\}$ $\subseteq$ $\NULLS(\vatoma\theta_\mathbf{i})$, 
	such that the following sequence of tuples of atoms:
	$\vatoma_{0}$, $\vatoma_{1}$, $\vatoma_{2}$, $\vatoma_{3}$, $\ldots$, $\vatoma_{s-1}$, $\vatoma_{s}$,
	%
	%
	represents how the two nulls
	$\SF{n}_{k_{j_0}}$ and $\SF{n}_{k_{j_\mathbf{i}}}$ converges onto the $\NULLS$-connected
	tuple $\vatoma\theta_\mathbf{i}$:
	\begin{align}
		\vatoma_{0}\,&=\,\LA\,\underbrace{\vatoma^{0}_{0}}_{\SF{n}_{m_0}},\hspace*{-0.15cm}
		                       \underbrace{\vatoma^{0}_{1}}_{\hspace*{-0.0cm}\leftarrow\,\,\,\SF{n}_{m_1}}
		                       \hspace*{-0.05cm},\hspace*{-0.15cm}
		                       \underbrace{\vatoma^{0}_{2}}_{\hspace*{-0.0cm}\leftarrow\,\,\,\SF{n}_{m_2}}
		                       \hspace*{-0.05cm},\hspace*{-0.15cm}
		                       ,\ldots,\hspace*{-0.1cm}
		                       \underbrace{\vatoma^{0}_{s}}_{\hspace*{-0.0cm}\leftarrow\,\,\,
		                       	                                                  \SF{n}_{m_s}}\RA
		                       \subseteq\CHASE^{j_\mathbf{i}}(D,\tgds);\nonumber\\
		\vatoma_{1}\,&=\,\LA\hspace*{-0.3cm}\underbrace{\vatoma^{1}_{0}\,\,,}_{\{\SF{n}_{m_0},\,\SF{n}_{m_1}\}}
		\hspace*{-0.0cm}
		\hspace*{-0.0cm}\underbrace{\vatoma^{1}_{2}}_{\hspace*{-0.0cm}\leftarrow\,\,\,\SF{n}_{m_2}}
							\hspace*{-0.05cm},\hspace*{-0.15cm}
							,\ldots,\hspace*{-0.1cm}
							\underbrace{\vatoma^{1}_{s}}_{\hspace*{-0.0cm}\leftarrow\,\,\,
								\SF{n}_{m_s}}\RA
							\subseteq\CHASE^{j_\mathbf{i}+t_1}(D,\tgds);\nonumber\\
		\vatoma_{2}\,&=\,\LA\hspace*{-0.3cm}\underbrace{\vatoma^{2}_{0}\,\,,}_{\{\SF{n}_{m_0},\,\SF{n}_{m_2}\}}\hspace*{-0.0cm}
		\hspace*{-0.0cm}
						\ldots,\hspace*{-0.1cm}
						\underbrace{\vatoma^{2}_{s}}_{\hspace*{-0.0cm}\leftarrow\,\,\,
							\SF{n}_{m_s}}\RA
						\subseteq\CHASE^{j_\mathbf{i}+t_2}(D,\tgds);\nonumber\\
		%
		%
		\vdots\hspace*{0.3cm}&\hspace*{1cm}\vdots\hspace*{1cm}\vdots\hspace*{1cm}&&\nonumber\\
		%
		%
		\vatoma_{s}\,&=\,\underbrace{\vatoma^s_{0}}_{\{\SF{n}_{m_0},\,\SF{n}_{m_s}\}}
		           \subseteq\CHASE^{j_\mathbf{i}+t_s}(D,\tgds),\mbox{ where:}\nonumber
	\end{align}

	\vspace*{-0.2cm}
	
	\noindent	
	\begin{enumerate}
		\item $\vatoma_{s}$ $=$ $\vatoma\theta_\mathbf{i}$;
		\item either: (a) $\SF{n}_{m_0}$ $=$ $\SF{n}_{k_{j_0}}$ and $\SF{n}_{m_s}$ $=$ $\SF{n}_{k_{j_\mathbf{i}}}$, or\\
		      \hspace*{0.9cm} (b) $\SF{n}_{m_0}$ $=$ $\SF{n}_{k_{j_\mathbf{i}}}$ and $\SF{n}_{m_s}$ $=$ $\SF{n}_{k_{j_0}}$\\
			  %
			  %
			  (i.e., either we move ``left-to-right" or ``right-to-left");	
		\item 
		      $\vatoma_{s}$ $=$ $\vatoma\theta_\mathbf{i}$, and 
		      $\vatoma^{p}_{0}$ and $\vatoma^{p}_{q}$ are tuples of atoms
		      \footnote{Here, the superscript ``$p$" on the tuples ``$\vatoma^{p}_{q}$" is simply used as a way to show 
		      	        tuples may possibly be distinct.};
		\item $\big|\{\SF{n}_{m_0},\SF{n}_{m_1},\SF{n}_{m_2},\SF{n}_{m_3},\ldots,\SF{n}_{m_{s-1}},
		      \SF{n}_{m_s}\}\big|$ $>$ $N'$ $=$ $N^N$;
        \item 
              $\SF{n}_{m_q}$ $\in$ $\NULLS(\vatoma^{p}_{q})$;
		\item 
		      $0$ $<$ $t_1$ $<$ $t_2$ $<$ $t_3$ $<$ $\ldots$ $<$ $t_{s-1}$ $<$ $t_{s}$;                      
		\item for each $n$ $\in$ $\{0,\ldots,s-1\}$, there exists a pair (``rule") 
		      $\LA B_n,\atomh_n\RA$ $\in$ $\tgds^{{\cal B}(\tgds)}$ and homomorphism
		      $\eta_n:$ $\VAR(B_n\cup \atomh_n)$ $\longrightarrow$ $\DOM(D)\cup\Gamma_\SF{N}$ such that:\label{seq_cond_5}
		      \begin{enumerate}
		      	\item $B_n\eta_n$ $\subseteq$ $\CHASE^{j_\mathbf{i}+t_n}(D,\tgds)$ and 
		      	      $\atomh_n\eta_n$ $\in$ $\CHASE^{j_\mathbf{i}+t_{n+1}}(D,\tgds)$;
		      	      \label{seq_cond_5_1}
		      	\item $\vatoma^n_{0}$ $\subseteq$ $B_n\eta_n$ and $\vatoma^{n+1}_{0}$ $=$ $\atomh_n\eta_n$. 
		      	      \label{seq_cond_5_2}		      	
		      \end{enumerate}
	\end{enumerate}
	Intuitively, the sequence of tuples (of atoms) $\vatoma_0$, $\vatoma_1$, $\vatoma_2$, $\vatoma_3$, 
	$\ldots$, $\vatoma_{s-1}$, $\vatoma_{s}$ above
	corresponds to the derivation of how the two nulls $\SF{n}_{k_{j_0}}$ and 
	$\SF{n}_{k_{j_\mathbf{i}}}$ actually ``converges" to the tuple $\vatoma\theta_\mathbf{i}$. 
	We note that because of the assumption (\ref{contra_condition}) above that the null 
	$\SF{n}_{k_{j_\mathbf{i}}}$ $\in$ $\big[\NULLS\big(\CHASE^{\,j_\mathbf{i}}(D,\tgds)\big)$
	$\mbox{\Large$\setminus$}$ $\NULLS\big(\CHASE^{\,j_{\mathbf{i}-1}}(D,\tgds)\big)\big]$	
	converges (i.e., appears in a $\NULLS$-connected tuple $\vatoma\theta$ $\subseteq$ $\CHASE(D,\tgds)$,
	           for some bijective homomorphism $\theta$)
	to all the other nulls in $\NULLS\big(\CHASE^{\,j_0}(D,\tgds)\big)$,		 
	it follows that the way $\SF{n}_{k_{j_0}}$ and 
	$\SF{n}_{k_{j_\mathbf{i}}}$ converges into the tuple $\vatoma\theta_\mathbf{i}$ must have come
	about from some of the sequence $\vatoma_0$, $\vatoma_1$, $\vatoma_2$, $\vatoma_3$, 
	$\ldots$, $\vatoma_{s-1}$, $\vatoma_{s}$ above and not through ``disjoint-atoms" in the bodies
	of rules in $\tgds$. Otherwise, the combination of nulls through a disjoint-atom in the rule bodies
	would make $\SF{n}_{k_{j_0}}$ and $\SF{n}_{k_{j_\mathbf{i}}}$ to be $\vatoma$-interchangeable since
	this would result in some homomorphism $\theta':$ $\NULLS(\vatoma\theta_\mathbf{i})$ $\longrightarrow$ $\Gamma_\SF{N}$
	such that $\theta'(\SF{n}_{k_{j_0}})$ $=$ $\theta'(\SF{n}_{k_{j_\mathbf{i}}})$
	and $\vatoma(\theta'\circ\theta_\mathbf{i})$ $\subseteq$ $\CHASE(D,\tgds)$. We emphasize that it is also
	from this fact that nulls must ``converge" into a single head atom in the chase derivation as well.	
	Then because of our assumption that
	$\mathbf{i}$ $>$ $N'$ $=$ $N^N$ $>$ $(\tgds^0)^{(\tgds^0)}$, 
	it follows that there must be a recursion in the way the rules in ``$\tgds^0$"
	was used in the generation of the sequence: $\vatoma_0$, $\vatoma_1$, $\vatoma_2$, $\vatoma_3$, 
		$\ldots$, $\vatoma_{s-1}$, $\vatoma_{s}$. Therefore, we get a contradiction because this 
	implies the existence of an RTC in $\tgds^{{\cal B}(\tgds)}$. 
	Indeed, it follows from the \emph{pigeonhole principle} that because 
	$\vect{i}$ $>$ $N'$ $=$ $N^N$ $\geq$ $(\tgds^0)^{(\tgds^0)}$
	(see again definition of $N$ in (\ref{N-val}) above), then even if all the rules
	in $\tgds^0$ participate in the converging of nulls simultaneously 
	(which can possibly cut-down the number of steps needed exponentially in making the nulls ``meet"
	 since more than one rule can be used in the process at the same time),
	then the fact that $\vect{i}$ $>$ $(\tgds^0)^{(\tgds^0)}$ implies
	we have a recursion in the way the rules in ``$\tgds^0$"
    was used in the generation of the sequence: $\vatoma_0$, $\vatoma_1$, $\vatoma_2$, $\vatoma_3$, 
	$\ldots$, $\vatoma_{s-1}$, $\vatoma_{s}$ above.
\end{proofsketch}

\vspace*{-0.3cm}


%
\begin{theorem}[\textbf{\textit{Finite controllability (FC) property}}]\label{fm_thm}
	For any database $D$, TGDs $\tgds$ $\in$ $\TG$ and BCQ $Q$, 
	$D$ $\cup$ $\tgds$ $\cup$ $\{\neg Q\}$ satisfies the finite model (FM) property.
\end{theorem}
\begin{proofsketch}
	Let $\vatoma$ $=$ $\atoma_1\ldots\atoma_l$ $=$ $\BD(x)$, for some
	$x$ $\in$ $\tgds\cup\{Q\}$. Then by Theorem \ref{bounded_nulls}, each of the null 
	$\SF{n}_j$ $\in$ $\big[\NULLS\big(\CHASE^{\,N+N'+k}(D,\tgds)\big)$ $\mbox{\Large $\setminus$}$
	$\NULLS\big(\CHASE^{\,N+N'}(D,\tgds)\big)\big]$ 
	is always $\vatoma$-interchangeable with some null $\SF{n}_i$ $\in$ $\NULLS\big(\CHASE^{\,N}(D,\tgds)\big)$.
	It then follows that $\CHASE(D,\tgds)$ can be represented by a finite number of nulls from which
	the finite model property also follows for the theory $D$ $\cup$ $\tgds$ $\cup$ $\{\neg Q\}$.
\end{proofsketch}

\vspace*{-0.3cm}

\begin{theorem}[\textbf{\textit{Comparison with other syntactic classes}}]\label{contains_others}
    For each class ${\cal C}$ $\in$ $\{\SC{wa}$, 
    $\SC{w-guarded}$, $\SC{wsj}$, $\SC{g-guarded}$, $\SC{shy}$, $\SC{tame}$,  $\SC{wr}\}$,
    we have that ${\cal C}$ $\subsetneq$ $\TG$. 
\end{theorem}
\begin{proofsketch2}
	On the contrary, assume that for some class ${\cal C}$ $\in$ 
	$\big\{\SC{wa}$, 
	      $\SC{w-guarded}$, $\SC{wsj}$, $\SC{g-guarded}$, 
	      $\SC{shy}$, $\SC{tame}$, $\SC{wr}\big\}$,
	we have that there exists a set of TGDs $\tgds$ s.t. $\tgds$ $\in$ ${\cal C}$ but
	$\tgds$ $\notin$ $\TG$. Then since $\tgds$ $\notin$ $\TG$, it follows from Definition 3
	that there exists an RTC: ${\cal T}$ $=$ $\big(\LA B,\atomh\RA$, $\LA\atoma,\atomb,\atomc\RA$, $\LA X,Z\RA$, $\atoma'\big)$
	%
	%
	%
    %
    (see Definition 2 and Figure 1),     
    %
	%
	where there does not exists an atom $\atomd$ $\in$ $B$ such that $\{X,Z\}$ $\subseteq$ $\VAR(\atomd)$.
	Then now let us consider the following cases:
	\begin{description}
		\item[Case 1:] ${\cal C}$ $=$ $\SC{wa}$: 
		    Then the fact that only the cyclically-affected
			variables are considered in Condition \textbf{4.)} of Definition 2 that 
			$\tgds$ $\in$ $\SC{wa}$.
			
		
			
		\item[Case 2:] ${\cal C}$ $=$ $\SC{w-guarded}$: 
			Then the fact that variables occurring in some cyclically-affected arguments
			are unguarded in ${\cal T}$ 
			(see Definition 3) contradicts the assumption that $\tgds$ $\in$ $\SC{w-guarded}$. 
			
		\item[Case 3:] ${\cal C}$ $=$ $\SC{wsj}$:
			Then the fact that each pair $\LA B,\atomh\RA$ $\in$ $\tgds^+$ and where we have from Definition 1
			that $\tgds^+$ also considers the ``TGD Expansion" in \cite{CaliGP12} used in the 
			identification of the \emph{sticky-join} class of TGDs, and because
			there exists marked variables $Y'$ $\in$ $\LINK(B,\,\atomd_{i},\,\atomd_{i+1})\mbox{\large $\setminus$}\{X,Z\}$ such that $Y'$ $\in$ $\VAR(\atoma')$ implies 
			all occurrences of variables in positions $\ARG(\atoma')\REST_{Y'}$
			in the atom $\atoma$ are in $\MVAR\big(\atoma,\vatomc,\atoma')$, 
			occurs in Condition \textbf{5.)}(b) of Definition 2 
			contradicts the assumption that $\tgds$ $\in$ $\SC{wsj}$.
			
		\item[Case 4:] ${\cal C}$ $=$ $\SC{g-guarded}$: 
		    Similarly to \textbf{Case 2} above, 
			the fact that the variables ``$X$" and ``$Z$" in ${\cal T}$ are such that
			$\{X,Z\}$ $\subseteq$ $\widehat{\VAR}(B)$ in Condition \textbf{4.)} of Definition 2,
			which are also the so-called ``glut-variables" \cite{KrotzschR11},
			then this contradicts that $\tgds$ $\in$ $\SC{g-guarded}$.
			%
			         				
		\item[Case 5:] ${\cal C}$ $=$ $\SC{shy}$:
			Then the fact that the linking variables $\LINK(\atomd_{i},\atomd_{i+1})$,
			for $i$ $\in$ $\{1,\ldots,m-1\}$, are non-empty in Condition \textbf{5.)}(b) 
			of Definition 2 contradicts that $\tgds$ $\in$ $\SC{shy}$.
			
		\item[Case 6:] ${\cal C}$ $=$ $\SC{tame}$:
			Since $\LA B,\atomh\RA$ $\in$ $\tgds^+$ and where $\tgds^+$ also consider the so-called 
			``tame reachability" in \cite{GottlobMP13} (see Definition 1), then it follows
			that some of the sticky rules ``fed" a guard atom of some guarded rule in $\tgds$ as resulting
			for the unification procedure in Definition 1. Then this contradicts that
			$\tgds$ $\in$ $\SC{tame}$.
			
			
		\item[Case 7:] ${\cal C}$ $=$ $\SC{wr}$:						
			Then the fact that the variables ``$X$" and ``$Z$" are unguarded in the RTC ${\cal T}$
			by Definition 3, and where each $\LINK(\atomd_i,\atomd_{i+1})$ are non-empty
			by Condition \textbf{5.)}(a) of Definition 2, implies that the
			``position graph" of $\tgds$ as described in \cite{wr-2012}, will have a cycle that passes
			through both an $m$-edge and an $s$-edge \big(i.e., the variables in $\LINK(\atomd_i,\atomd_{i+1})$\big). 
			Therefore, this contradicts the assumption that $\tgds$ $\in$ $\SC{wr}$. $\Box$
	\end{description}
\end{proofsketch2}

\vspace*{-0.1cm}

\begin{theorem}[\textbf{\textit{BCQ-Ans complexity}}]\label{BCQ-Ans_complexity}
	The BCQ-Ans combined complexity problem under the class $\TG$ is in $4$-\SC{ExpTime}
	but is $3$-\SC{ExpTime}-hard in general.		
\end{theorem}
\begin{proofsketch} (``\emph{upper-bound}") Let $D$ be a database, $\tgds$ $\in$ $\TG$ a set of TGDs in $\TG$ and $Q$ a BCQ. 
	Then with the numbers $N'$ $=$ $N^N$ and $N$ as defined in the proof of Theorem \ref{bounded_nulls}
	\big(i.e., see (\ref{N-val})\big), then we get
	that it is sufficient to only consider chase depths ``$\CHASE^{\,K}(D,\tgds)$" where $K$ is of the order
	$O(2^{2^{p(n)}})$ \big(i.e., \emph{doubly-exponential} to $p(n)$\big), and $p(n)$ is a polynomial to 
	$|D|$, $|\tgds|$ and $|Q|$. It then follows from Theorem 5 in \cite{ZhangZY15} that it is
	$\big(K+2\big)$-\SC{ExpTime} to check if $\CHASE^{\,K}(D,\tgds)$ $\models$ $Q$.

    (``\emph{lower-bound}") Since it follows from \cite{KrotzschR11} that BCQ-Ans under the $\SC{g-guarded}$ class
	is $3$-\SC{ExpTime}-complete, then because we have from Theorem \ref{contains_others} that our new $\TG$ class
	also contains $\SC{g-guarded}$, then if follows that BCQ-Ans under $\TG$ is at least $3$-\SC{ExpTime}-hard.
\end{proofsketch}

\vspace*{-0.6cm}

\section{Concluding Remarks}

In this paper, we have introduced a new class of TGDs called \emph{triangularly-guarded} TGDs 
($\TG$), for which BCQ-Ans is decidable as well as having the FC property (Theorem 
\ref{fm_thm}). We further showed that $\TG$ strictly contains the current main syntactic classes: 
\SC{wa}, \SC{w-guarded}, \SC{wsj}, \SC{g-guarded},
\SC{shy}, \SC{tame} and \SC{wr} (Theorem \ref{contains_others}), which, to the best of our knowledge,
provides a unified representation of those aforementioned TGD classes. 
Since our new $\TG$ class of TGDs satisfies the FC property as
well as strictly contains all the other important classes mentioned above, 
this work also provides an important and unifying explanation 
why those other important classes satisfies the FC property
and thus, are decidable.




\bibliographystyle{aaai}
\bibliography{TG_TGDs_KR-2018}

\end{document}